\RequirePackage{fix-cm}
\documentclass[smallextended]{svjour3}       % onecolumn (second format)
\smartqed  % flush right qed marks, e.g. at end of proof
\usepackage{graphicx}
\begin{document}

\title{Spectrum of nonstrange singly charmed baryons in the constituent quark model%\thanks{Grants or other notes
}


\author{Keval Gandhi         \and
        Zalak Shah  \and
        Ajay Kumar Rai %etc.
}


\institute{Keval Gandhi \at
              Department of Applied Physics, Sardar Vallabhbhai National
Institute of Technology, Surat$-$395007, Gujarat, India.\\
              \email{keval.physics@yahoo.com}           %  \\
           \and
           Zalak Shah \at
              Department of Applied Physics, Sardar Vallabhbhai National
Institute of Technology, Surat$-$395007, Gujarat, India.\\
\email{zalak.physics@gmail.com} 
\and
           Ajay Kumar Rai \at
              Department of Applied Physics, Sardar Vallabhbhai National
Institute of Technology, Surat$-$395007, Gujarat, India.\\
\email{raiajayk@gmail.com}
}

\date{Received: date / Accepted: date}

\maketitle

\begin{abstract}
We study the masses of radial and orbital excited states of nonstrange singly charmed baryons in the framework of hypercentral Constituent Quark Model (hCQM). To obtain the mass spectra, the Coulomb plus screened potential is employed with the first order correction, which gives a relativistic effect of order $\cal{O}$$(1/m)$. The spin-spin, spin-orbit and tensor interactions are included (perturbatively as a spin dependent potential) in order to generate the splitting in mass spectra. We compare our computed mass spectra of nonstrange singly charmed baryons with the other theoretical predictions as well as with the experimental observations. We construct the Regge trajectories of these baryons in the $(J, M^2)$ plane. Further, we analyze the strong one pion decay rates for $S$, $P$ and the $D$-wave transitions in the framework of Heavy Hadron Chiral Perturbation Theory (HHChPT). Moreover, the electromagnetic properties like magnetic moments, transition magnetic moments and the radiative decay widths are determined for the ground state of these baryons in the constituent quark model. 
\end{abstract}

\section{Introduction}
\label{intro}

The nonstrange singly charmed baryons belong to $\Lambda{_{c}^+}$ and $\Sigma_c$ families, which are classified into SU(3) flavor representation of antisymmetric antitriplet and symmetric sextet group respectively. Their experimental evidences are continuously in progress from the past few years. The latest Review of Particle Physics (RPP) by Particle Data Group (PDG) \cite{Tanabashi2018} present the seven states of $\Lambda{_{c}^+}$ baryon: $\Lambda{_{c}(2286)^+}$, $\Lambda{_{c}(2595)^+}$,  $\Lambda{_{c}(2625)^+}$, $\Lambda{_{c}(2765)^+}$, $\Lambda{_{c}(2860)^+}$, $\Lambda{_{c}(2880)^+}$, $\Lambda{_{c}(2940)^+}$ and the three states of isotriplet $\Sigma_c$  baryons:  $\Sigma{_{c}(2455)^{++,+,0}}$, $\Sigma{_{c}(2520)^{++,+,0}}$, $\Sigma{_{c}(2800)^{++,+,0}}$ (see in Table \ref{tab:1}).  For that the various experimental groups FOCUS, CLEO, \textit{BABAR}, CDF and Belle have provided the masses as well as other properties of these baryons. Recently, the LHCb Collaboration \cite{Aaij2017} has given precise measurements of the masses and the strong decay widths of $\Lambda{_{c}^+}$ and the isotriplet $\Sigma_c$ baryons with the statistical and the systematic uncertainties. The more experimental informations are available in review articles \cite{Chen2017,Sonnenschein2018}. Moreover, the current projects LHCb and Belle II and the future experimental facilities J-PARC, $\overline{\mbox{\sffamily P}}${\sffamily ANDA} \cite{Barucca2019,Singh(all)} are expected to provide further information regarding singly charmed baryons.\\

The LHCb Collaboration \cite{Aaij2017} assigned the $J^P$ ($J$ is the total spin and $P$ is parity) value of $\Lambda{_{c}(2880)^+}$ and $\Lambda{_{c}(2940)^+}$ states as ${\frac{5}{2}}^+$ and ${\frac{3}{2}}^-$ respectively. They also measured a state $\Lambda{_{c}(2860)^+}$ with $J^P = {\frac{3}{2}}^+$. Except $\Lambda{_{c}(2765)^+}$, the spin-parity of all observed $\Lambda{_{c}^+}$ baryons have been confirmed as shown in Table \ref{tab:1}. On the other hand the $J^P$ value of excited isotriplet $\Sigma_c$ baryons have not been confirmed yet. The identifications and decay properties of new states of these baryons makes this study challenging. It is interesting to look back to the theory and the phenomenological study to see where the new predictions lie. Till date, the mass spectra of singly charmed baryons have been studied in various potential models using different approaches: a Quantum Chromodynamics (QCD) based quark model \cite{Copley1979}, relativistic quark potential model \cite{Capstick1986}, relativistic quark-diquark picture \cite{Ebert2008,Ebert2011}, QCD sum rule \cite{Chen2015}, Faddeev method \cite{Valcarce2008} the flux tube model \cite{Bing2009}, the quasi-linear Regge trajectory ansatz \cite{Guo2008}, the non-relativistic constituent quark model \cite{Roberts2008,Yoshida2015,Shah2016cpc,Shah2016epja,Gandhi2019}, heavy quark limit in the one-boson-exchange potential \cite{Yamaguchi2015}, the heavy quark-light quark cluster picture \cite{BChen2017}, lattice QCD study \cite{Mathur2002,Padmanath2013,Rubio2015} etc..\\

\begin{table}
%\begin{center}
\caption{Mass, width and $J^P$ value of the nonstrange singly charmed baryons from PDG \cite{Tanabashi2018}.}
\label{tab:1}
%\scalebox{1.0}{
\begin{tabular}{ccccccccccccccccccc}
\hline\noalign{\smallskip}
Resonance & Mass (MeV) & Width (MeV) & $J^P$ \\
\noalign{\smallskip}\hline\noalign{\smallskip}
$\Lambda{_{c}^+}$ & 2286.46 $\pm$ 0.14 & $-$ & ${\frac{1}{2}}^+$ \\
$\Lambda{_{c}(2595)^+}$ & 2592.25 $\pm$ 0.28 & 2.59 $\pm$ 0.30 $\pm$ 0.47 & ${\frac{1}{2}}^-$\\
$\Lambda{_{c}(2625)^+}$ & 2628.11 $\pm$ 0.19 & $<$0.97 & ${\frac{3}{2}}^-$\\
$\Lambda{_{c}(2765)^+}$ & 2766.6 $\pm$ 2.4 & 50 & $?^?$ \\
$\Lambda{_{c}(2860)^+}$ & 2856.1${_{-1.7}^{+2.0}}$ $\pm$ 0.5${_{-5.6}^{+1.1}}$ & 67.6${_{-8.1}^{+10.1}}$ $\pm$ 1.4${_{-20.0}^{+5.9}}$ & ${\frac{3}{2}}^+$ \\
$\Lambda{_{c}(2880)^+}$ & 2881.62 $\pm$ 0.24 & 5.6${_{-0.6}^{+0.8}}$ & ${\frac{5}{2}}^+$\\
$\Lambda{_{c}(2940)^+}$ & 2939.6${_{-1.5}^{+1.3}}$ & 20${_{-5}^{+6}}$ & ${\frac{3}{2}}^-$ \\
$\Sigma{_{c}(2455)^{++}}$ & 2453.97 $\pm$ 0.14 & 1.89${_{-0.18}^{+0.09}}$ & ${\frac{1}{2}}^+$ \\
$\Sigma{_{c}(2455)^{+}}$ & 2452.9 $\pm$ 0.4 & $<$4.6 & ${\frac{1}{2}}^+$ \\
$\Sigma{_{c}(2455)^{0}}$ & 2453.75 $\pm$ 0.14 & 1.83${_{-0.19}^{+0.11}}$ & ${\frac{1}{2}}^+$\\
$\Sigma{_{c}(2520)^{++}}$ & 2518.41${_{-0.19}^{+0.21}}$ & 14.78${_{-0.40}^{+0.30}}$ & ${\frac{3}{2}}^+$ \\
$\Sigma{_{c}(2520)^{+}}$ & 2517.5 $\pm$ 2.3 & $<$17 & ${\frac{3}{2}}^+$ \\
$\Sigma{_{c}(2520)^{0}}$ & 2518.48 $\pm$ 0.20 & 15.3${_{-0.5}^{+0.4}}$ & ${\frac{3}{2}}^+$ \\
$\Sigma{_{c}(2800)^{++}}$ & 2801${_{-6}^{+4}}$ & $75{_{-13-11}^{+18+12}}$ & $?^?$ \\
$\Sigma{_{c}(2800)^{+}}$ & 2792${_{-5}^{+14}}$ & $62{_{-23-38}^{+37+52}}$ & $?^?$ \\
$\Sigma{_{c}(2800)^{0}}$ & 2806${_{-7}^{+5}}$ & $72{_{-15}^{+22}}$ & $?^?$ \\
\noalign{\smallskip}\hline
\end{tabular}
%\end{center}
\end{table}

In order to improve the understanding of quark dynamics and their confinement mechanism inside the baryons, the study of singly charmed baryons containing one heavy quark and two light quarks is an important tool. It can provide some qualitative informations about the chiral symmetry breaking and the heavy quark symmetry. The singly charmed baryons mainly decay via strong interactions and it will be dominant over the electromagnetic or weak decay processes. So far any electromagnetic observation of the singly charmed baryons are not found experimentally. The strong decays and the electromagnetic behavior of the charmed baryons have been studied by several methods and they are: the Heavy Hadron Chiral Perturbation Theory (HHChPT) \cite{Cheng1993,Yan1992,Cho1994,Cheng1997,Pirjol1997,Cheng2007,Cheng2015,Cheng20151,Gandhi2018,Jiang2015}, the chiral structure model \cite{Meng2018,Sharma2010,Kawakami2018}, the chiral soliton model \cite{Kim2018}, the pion mean-field approach \cite{HCKim2019}, the chiral perturbation theory \cite{Shi2018,Wang2019}, the relativistic constituent three-quark model \cite{Ivanov1999,Faessler2006}, large $N_c$ limit \cite{Guralnik1993,Jenkins1993,Yang2018}, the light front quark model \cite{Tawfiq1998}, the QCD sum rule \cite{Aliev2009,Aliev20091,Aliev2012,Wang2017a,Wang2018b}, the $^3P_0$ model \cite{Chen2007,Lu2018}, a constituent quark model \cite{Wang2018,Yao2018}, the bag model \cite{Bernotas2013}, the non-relativistic approach \cite{Wang2017,Albertus2005,Patel2008,Dhir2009,Shah2019,Majethiya2009}, the lattice QCD \cite{Detmold2012,Can2014,Bahtiyar2017} etc..\\

In this paper, the masses of radial and orbital excited states of nonstrange singly charmed baryons are calculated. For that, we employ the hypercentral Constituent Quark Model (hCQM) in which Coulomb plus screening potential \textbf{is used} with the first order correction. The obtained mass spectra are presented corresponding to $n = 1, 2, 3, 4, 5$ (n is the principal quantum number) with orbital quantum number $L = 0, 1, 2, 3$. Such masses are used to draw Regge trajectories in the $(J, M^2)$ plane.\\

This paper is organized as follows:  After the introduction, in section 2 we present details of the hypercentral Constituent Quark Model (hCQM) and discuss the potential model. In section 3 we analyze the mass spectra and construct the Regge trajectories. In section 4 the strong one pion decay rates of $\Lambda{_{c}(2765)^+}$, $\Sigma_c(2455)$, $\Sigma_c(2520)$ and $\Sigma_c(2800)$ baryons are calculated in HHChPT, and also the electromagnetic properties such as the magnetic moments, transition magnetic moments and the radiative decays for $L = 0$ are studied in the constituent quark model. At last, we summarize our present work in section 5. 

\section{Methodology}
\label{sec:2}

\noindent The spectroscopy of light and heavy flavor baryons are usually studied in relativistic (or non-relativistic) approach of quantum mechanics. In this section, we introduce the non-relativistic treatment in the framework of hypercentral Constituent Quark Model  (hCQM). Such a model is well established and have been used to determine the properties of light, heavy-light and heavy-heavy flavored baryons (see in Refs. \cite{Shah2016cpc,Shah2016epja,Shah2019,Ghalenovia2014,Santopinto2005}). It has a spin-independent term with confining potential, gives a confining effect at a long range quark separations. The relative Jacobi coordinates ($\vec{\rho}$ and $\vec{\lambda}$) are employed to see the dynamics of three quarks system. We express the Hamiltonian as \cite{Richard1992,Giannini2015},

\begin{equation}
\label{eqn:1} 
H=\frac{P{_{x}^2}}{2m} + V(x)
\end{equation} 

\noindent where $x$ is the six-dimensional hypercentral coordinate and $ m = \frac{2m_\rho m_\lambda}{m_\rho+m_\lambda} $ gives the reduced mass of the baryonic system. Here, the coordinates $\vec{\rho}$ and $\vec{\lambda}$ are, 

\begin{equation}
\label{eqn:2} 
\vec{\rho}=\frac{1}{\sqrt2} (\vec{r_1}-\vec{r_2}) \hspace{0.5cm} and \hspace{0.5cm} \vec{\lambda}=\frac{\vec{r_1}+\vec{r_2}-2\vec{r_3}} {\sqrt {6}};
\end{equation}

\noindent the relative Jacobi coordinates, that describe the relevant degrees of freedom for the dynamics of three constituent quarks. Here, $\vec{r_i}$ ${(i=1, 2, 3)}$ represents the $i^{th}$ coordinate of the constituent quarks. The reduced mass of these coordinates are \cite{Bijker2000},

\begin{equation}
\label{eqn:3} 
m_\rho = \frac{2m_1m_2}{(m_1+m_2)} \hspace{0.5cm} and \hspace{0.5cm} m_\lambda=\frac{2m_3(m{_{1}^2}+m{_{1}^2}+m_1m_2)}{(m_1+m_2)(m_1+m_2+m_3)}.
\end{equation}

For the $\Lambda_{c}^+$ baryon the constituent quarks are $u$, $d$ and $c$ and the $\Sigma_c$ baryon has three isospin states with a different quark constitutions: $\Sigma{_{c}^{++}}$ $(uuc)$, $\Sigma{_{c}^{+}}$ $(udc)$ and $\Sigma{_{c}^{0}}$ $(ddc)$. To calculate the masses of these isotriplet $\Sigma_c$ baryons separately, we consider different constituent quark masses (taken from Ref. \cite{Shah2016epja}) of the light quarks $(u$ and $d)$, which are $ m_u = 0.338 $ GeV, $ m_d = 0.350 $ GeV and for the charm quark $(c)$, $ m_c = 1.275 $ GeV.\\

The hyperspherical coordinates, hyperradias $(x)$ and hyperangle $(\xi)$, are defined in the form of Jacobi coordinates ($\vec{\rho}$ and $\vec{\lambda}$) as, $ x=\sqrt{\rho^2+\lambda^2} $ and $ \xi= arctan \left( \frac{\rho^2}{\lambda^2} \right) $. In the center-of-mass frame, an expression of kinetic energy operator $ \frac{P{_{x}^2}}{2m} $ (appear in the Eq. (\ref{eqn:1})) for a three quarks system is,

\begin{equation}
\label{eqn:4} 
\frac{P{_{x}^2}}{2m}=-\frac{\hbar^2}{2m}(\triangle_\rho + \triangle_\lambda)=-\frac{\hbar^2}{2m} \left(\frac{\partial^2}{\partial x^2}+\frac{5}{x}\frac{\partial}{\partial x}+\frac{L^2(\Omega)}{x^2}\right)
\end{equation}

\noindent where $ L^2(\Omega)=L^2(\Omega_\rho, \Omega_\lambda, \xi ) $ represents the quadratic Casimir operator in the six-dimensional rotational group $O(6)$ and its eigenfunctions are the hyperspherical harmonics $ Y_{[\gamma]}l_{\rho}l_{\lambda} (\Omega_\rho, \Omega_\lambda, \xi ) $ satisfying the eigenvalue relation,

\begin{equation}
\label{eqn:5} 
L^2 Y_{[\gamma]}l_{\rho}l_{\lambda} (\Omega_\rho, \Omega_\lambda, \xi) = -\gamma(\gamma+4)Y_{[\gamma]}l_{\rho}l_{\lambda} (\Omega_\rho, \Omega_\lambda, \xi).
\end{equation}

Here, $\Omega_\rho$ and $\Omega_\lambda$ are the angles of the hyperspherical coordinates. The total angular momentum is $ \vec{L}=\vec{L_\rho}+\vec{L_\lambda} $ , and the angular momentum associated with the Jacobi coordinates $ \vec{\rho} $ and $ \vec{\lambda} $ are, $l_\rho$ and $l_\lambda$, respectively. The eigenvalues of $L^2$ are given by $-\gamma(\gamma+4)$, where $\gamma = 2n + \vec{\rho} + \vec{\lambda} $ represents the grand angular momentum quantum number with the non-negative integer value ($n$).

\subsection{The Potential Model}

\begin{figure*}
\resizebox{1.0\textwidth}{!}{%
  \includegraphics{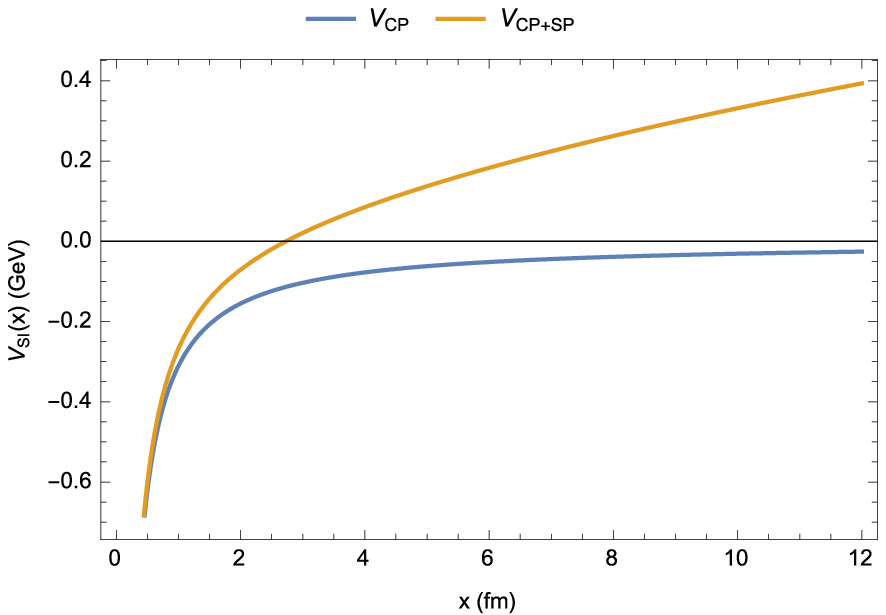}
  \includegraphics{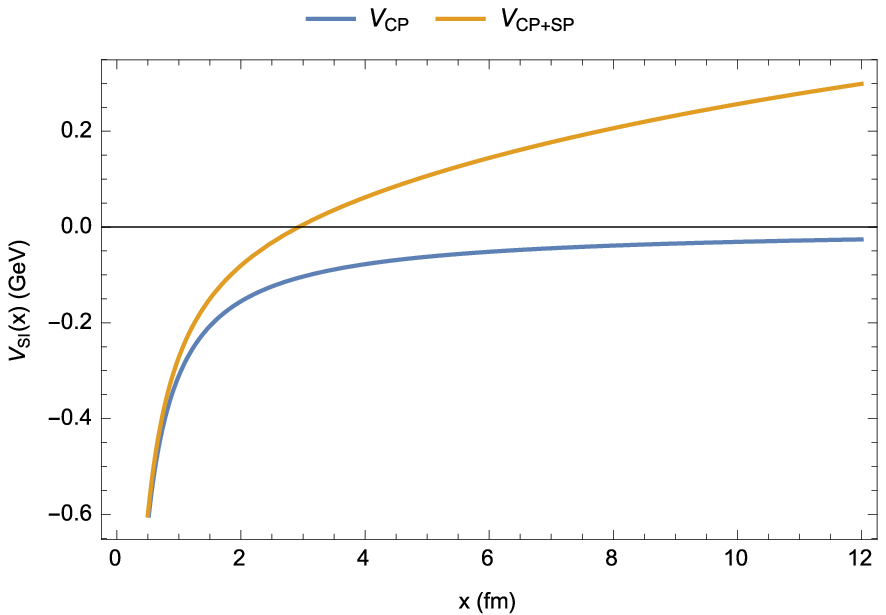}
}
\caption{The nature of non-relativistic spin independent part of QCD potential $(V_{SI}(x))$, $\Lambda{_{c}^+}$ baryon (left) and $\Sigma{_{c}^+}$ baryon (right). $V_{SI}(x)$ is changing with respect to inter-quark separation $(x)$ for $\alpha_s = 0.46 $. The blue line indicates the short range Coulomb interaction ($V_{CP}$) and the added long range interaction potential say Coulomb plus screened potential ($V_{CP+SP}$) is presented by orange line.}
\label{Fig:1}       % Give a unique label
\end{figure*}

The non-relativistic interaction potential $V(x)$ classified into spin-independent $ V_{SI}(x) $ and spin-dependent $ V_{SD}(x) $ part of the potential. The effective spin-independent static potential $ V_{SI}(x) $ is simply the sum of Lorentz vector (Coulomb) $V_V(x)$ and Lorentz scalar (confining) $V_S(x)$ terms. In the present study we used the first order correction (say $V_I(x)$) in $ V_{SI}(x) $ \cite{Koma2006}, i.e.,

\begin{equation}
\label{eqn:6} 
V_{SI}(x)=V_V(x)+V_S(x)+V_I(x).
\end{equation}

Here, $V_V(x)$ is the non-relativistic QCD potential between two quarks $q_1$ and $q_2$ in a baryon $(q_1q_2q_3)$, which is written in a Coulombic form by considering color wavefunction as, 

\begin{equation}
\label{eqn:7} 
V_V(x) = -\frac{2}{3}\frac{\alpha_s}{x}, 
\end{equation}

\noindent where $x$ is the inter-quark separation and the parameter $\alpha_s$ represents the strong running coupling constant

\begin{equation}
\label{eqn:8} 
\alpha_s = \frac{\alpha_s(\mu_0)}{1+\left( \frac{33-2n_f}{12\pi}\right)\alpha_s(\mu_0)ln\left(\frac{m_1+m_2+m_3}{\mu_{0}}\right)}.
\end{equation}

In the above expression, we consider $\alpha_s(\mu_0=1GeV) = 0.6$ and $n_f$ is the number of active quark flavors contributing effectively in quark-gloun loops. The quantity $(33-2n_f)$ must be greater than zero. Therefore, $n_f$ will never larger than six. In the present calculation we take $n_f=3$. So we have an approximately $\alpha_s = 0.46$ for both $\Lambda_{c}^+$ and $\Sigma_c$ baryons. We choose screened potential as a scalar potential,

\begin{equation}
V_{S}(x)= a \left(\frac{1-e^{-\mu x}}{\mu}\right),
\end{equation}

\noindent where $a$ is the string tension and the constant $ \mu $ occurring in screened potential is the screening factor. For $ x \ll \frac{1}{\mu} $, the screened potential is behaving like a linear potential $ax$ and for $ x \gg \frac{1}{\mu} $ it becomes a constant $ \frac{a}{\mu} $. So it is interesting to study the mass spectroscopy with screened potential which gives the masses of the higher excited states lesser compared to the linear potential \cite{JZWang2019,BQLi20091}. Here, we set $\mu = 0.04$ GeV. Fig. (\ref{Fig:1}) shows the behavior of spin-independent Coulomb plus screened potential $(V_{CP+SP})$ for $\Lambda{_{c}^+}$ and $\Sigma{_{c}^+}$ baryons.\\

\begin{table}
\caption{\label{tab:2}Predicted masses of the radial and the orbital excited states of $\Lambda{_{c}^+}$ baryon (in GeV).}
\scalebox{1}{
     % Give a unique label
\begin{tabular}{cccccccccccccc}
\hline\noalign{\smallskip}
State & $J^P$& Present  & \cite{Ebert2008} & \cite{Ebert2011} & \cite{Chen2015} & \cite{Yoshida2015} & \cite{Yamaguchi2015} & \cite{BChen2017} & PDG \cite{Tanabashi2018} \\
\noalign{\smallskip}\hline\noalign{\smallskip}
$1S$ & ${\frac{1}{2}}^+$ & 2.286 & 2.297 & 2.286 & 2.286 & 2.285 & 2.298 & 2.286 & 2.28646$\pm$0.00014 \\
$2S$ & ${\frac{1}{2}}^+$ & 2.785 & 2.772 & 2.769 & 2.766 & 2.857 & 2.791 & 2.772 & 2.7666$\pm$0.00024 \\
$3S$ & ${\frac{1}{2}}^+$ & 3.138 & 3.150 & 3.130 & 3.112 & 3.123 & 2.983 & 3.116 \\
$4S$ & ${\frac{1}{2}}^+$ & 3.454 & & 3.437 & 3.397 & & 3.154\\
$5S$ & ${\frac{1}{2}}^+$ & 3.742 & & 3.715 &  \\
\noalign{\smallskip}\hline\noalign{\smallskip}
$1P$ & ${\frac{1}{2}}^-$  & 2.573 & 2.598 & 2.598 & 2.591 & 2.628 & 2.625 & 2.614 & 2.59225$\pm$0.00028\\
$1P$ & ${\frac{3}{2}}^-$  & 2.568 & 2.628 & 2.627 & 2.629 & 2.630 & 2.816 & 2.639 & 2.62811$\pm$0.00019\\
\noalign{\smallskip}\hline\noalign{\smallskip}
$2P$ & ${\frac{1}{2}}^-$  & 2.978  & 3.017 & 2.983 & 2.989 & 2.890 & & 2.980  \\
$2P$ & ${\frac{3}{2}}^-$ & 2.970 & 3.034 & 3.005 & 3.000 & 2.917 & & 3.004 & 2.9396${_{-0.0015}^{+0.0014}}$ \\
\noalign{\smallskip}\hline\noalign{\smallskip}
$3P$ & ${\frac{1}{2}}^-$  & 3.392 & & 3.303 & 3.296 & 2.933 \\
$3P$ & ${\frac{3}{2}}^-$ & 3.384 & & 3.322 & 3.301 & 2.956 \\
\noalign{\smallskip}\hline\noalign{\smallskip}
$4P$ & ${\frac{1}{2}}^-$ & 3.813 & & 3.588 & \\
$4P$ & ${\frac{3}{2}}^-$ & 3.804 & & 3.606 \\
\noalign{\smallskip}\hline\noalign{\smallskip}
$5P$ & ${\frac{1}{2}}^-$ & 4.240  & & 3.852 \\
$5P$ & ${\frac{3}{2}}^-$ & 4.230 & & 3.869 \\
\noalign{\smallskip}\hline\noalign{\smallskip}
$1D$ & ${\frac{3}{2}}^+$ & 2.876 & 2.874 & 2.874 & 2.857 & 2.920 & 3.12 & 2.843 & 2.8561$\pm$0.0005\\
$1D$ & ${\frac{5}{2}}^+$ & 2.865 & 2.883 & 2.880 & 2.879 & 2.922 & 3.125 & 2.851 & 2.88162$\pm$0.00024\\
\noalign{\smallskip}\hline\noalign{\smallskip}
$2D$ & ${\frac{3}{2}}^+$ & 3.256 & 3.262 & 3.189 & 3.188 & 3.175 & 3.194\\
$2D$ & ${\frac{5}{2}}^+$ & 3.244 & 3.268 & 3.209 & 3.198 & 3.202 & 3.194\\
\noalign{\smallskip}\hline\noalign{\smallskip}
$3D$ & ${\frac{3}{2}}^+$ & 3.639 & & 3.480 &  & 3.191\\
$3D$ & ${\frac{5}{2}}^+$ & 3.627 & & 3.500 &  & 3.230\\
\noalign{\smallskip}\hline\noalign{\smallskip}
$4D$ & ${\frac{3}{2}}^+$ & 4.024 & & 3.747\\
$4D$ & ${\frac{5}{2}}^+$ & 4.012 & & 3.767\\
\noalign{\smallskip}\hline\noalign{\smallskip}
$5D$ & ${\frac{3}{2}}^+$ & 4.410 & \\
$5D$ & ${\frac{5}{2}}^+$ & 4.397 & \\
\noalign{\smallskip}\hline\noalign{\smallskip}
$1F$ & ${\frac{5}{2}}^-$ & 3.152 & 3.061 & 3.097 & 3.075 & 2.960 & 3.092 \\
$1F$ & ${\frac{7}{2}}^-$ & 3.136 & 3.057 & 3.078 & 3.092 && 3.128 \\
\noalign{\smallskip}\hline\noalign{\smallskip}
$2F$ & ${\frac{5}{2}}^-$ & 3.517 & & 3.375 & & 3.444\\
$2F$ & ${\frac{7}{2}}^-$ & 3.500 & & 3.393 \\
\noalign{\smallskip}\hline\noalign{\smallskip}
$3F$ & ${\frac{5}{2}}^-$ & 3.880 & & 3.646 & & 3.491\\
$3F$ & ${\frac{7}{2}}^-$ & 3.865 & & 3.667\\
\noalign{\smallskip}\hline\noalign{\smallskip}
$4F$ & ${\frac{5}{2}}^-$ & 4.244 & & 3.900\\
$4F$ & ${\frac{7}{2}}^-$ & 4.228 & & 3.922\\
\noalign{\smallskip}\hline\noalign{\smallskip}
$5F$ & ${\frac{5}{2}}^-$ & 4.604 &\\
$5F$ & ${\frac{7}{2}}^-$ & 4.590 &\\
\noalign{\smallskip}\hline
\end{tabular}}
% Or use
 % with the correct table height
\end{table}

\begin{table}
\caption{\label{tab:3}Predicted masses of the radial excited states of isotriplet $\Sigma_c$ baryons (in GeV).}
\scalebox{1}{
% Give a unique label
\begin{tabular}{ccccccccccccc}%\begin{tabular}{lllllllll}
\hline\noalign{\smallskip}
Particle&State & $J^{P}$ & Present & \cite{Capstick1986} & \cite{Ebert2008} & \cite{Ebert2011} & \cite{Yoshida2015} & \cite{Yamaguchi2015} & \cite{BChen2017} & PDG   \cite{Tanabashi2018} \\
\noalign{\smallskip}\hline\noalign{\smallskip}
$\Sigma{_{c}^{++}}$&1S & $\frac{1}{2}^{+}$ & 2.447 & & & & & & & 2.45397 $\pm$ 0.00014\\
&2S	& & 2.904 \\
&3S	& & 3.260 \\
&4S	& & 3.588 \\
&5S	& & 3.896 \\
\noalign{\smallskip}
\cline{2-11}
\noalign{\smallskip}
&1S&$\frac{3}{2}^{+}$& 2.512 & &  &  & & & &  2.51841${_{-0.00019}^{+0.00021}}$ \\
&2S& & 2.943  \\
&3S& & 3.282  \\
&4S& & 3.602  \\
&5S& & 3.905 \\
\noalign{\smallskip}\hline\noalign{\smallskip}
$\Sigma{_{c}^+}$&1S & $\frac{1}{2}^{+}$ & 2.456 & 2.440 & 2.439 & 2.443 & 2.460 & 2.455 & 2.456 & 2.4529 $\pm$ 0.0004\\
&2S	& & 2.912  & 2.890 & 2.864 & 2.901 & 3.029 & 2.958 & 2.850\\
&3S	& & 3.266  & 3.035 & & 3.271 & 3.103 & 3.115 & 3.091\\
&4S	& & 3.593 \\
&5S	& & 3.900 \\
\noalign{\smallskip}
\cline{2-11}
\noalign{\smallskip}
&1S&$\frac{3}{2}^{+}$& 2.521 & 2.495 & 2.518 & 2.519 & 2.523 & 2.519 & 2.515 & 2.5175 $\pm$ 0.0023 \\
&2S& & 2.951 & 2.985 & 2.912 & 2.936 & 3.065 & 2.876 & 2.995 &  \\
&3S& & 3.288 & 3.200 & & 3.109 & 3.094 & 3.116 & 3.091\\
&4S& & 3.606 \\
&5S& & 3.909 \\
\noalign{\smallskip}\hline\noalign{\smallskip}
$ \Sigma{_{c}^0}$ &1S &$\frac{1}{2}^{+}$ & 2.466 &  & & & & & & 2.45375 $\pm$ 0.00014 \\
& 2S	& & 2.919 \\
& 3S	& & 3.272\\
& 4S	& & 3.598	\\
& 5S	& & 3.904 \\
\noalign{\smallskip}
\cline{2-11}
\noalign{\smallskip}
&1S&$\frac{3}{2}^{+}$& 2.530 & & & & & & & 2.51848 $\pm$ 0.00020\\
&2S& & 2.958\\
&3S& & 3.294\\
&4S	& & 3.611\\
&5S& & 3.913\\
\noalign{\smallskip}\hline
\end{tabular}}
% Or use
 % with the correct table height
\end{table}

\begin{table}
\caption{Predicted masses of the orbital excited states of $\Sigma{_{c}^{++}}$ baryon (in GeV).}
\label{tab:4}  
     % Give a unique label
\begin{tabular}{ccccccccccccc}%\begin{tabular}{lllllllll}
\hline\noalign{\smallskip}
State & $J^P$ & Present & PDG \cite{Tanabashi2018} & & State & $J^P$ & Present & PDG \cite{Tanabashi2018} \\
\noalign{\smallskip}\noalign{\smallskip}
\cline{1-4}
\cline{6-9}
\noalign{\smallskip}\noalign{\smallskip}
1P & $\frac{1}{2}^{-}$ &	2.796	 & 2.801${_{-0.006}^{+0.004}}$ & & 4D & $\frac{3}{2}^{+}$ & 3.776\\
& $\frac{3}{2}^{-}$  & 	2.785 & & & & $\frac{5}{2}^{+}$ & 	3.768\\
& $\frac{1}{2}^{-}$  & 	2.802 & & & & $\frac{1}{2}^{+}$ & 	3.784\\
& $\frac{3}{2}^{-}$  & 	2.791 & & & & $\frac{3}{2}^{+}$ & 	3.779\\
& $\frac{5}{2}^{-}$  & 	2.776 & & & & $\frac{5}{2}^{+}$ & 	3.771\\
&&&&&& $\frac{7}{2}^{+}$ & 	3.761\\
\noalign{\smallskip}
\cline{1-4}
\cline{6-9}
\noalign{\smallskip}
2P& $\frac{1}{2}^{-}$ &	3.092 & & & 5D & $\frac{3}{2}^{+}$ & 3.997\\
& $\frac{3}{2}^{-}$  & 	3.082 & & & & $\frac{5}{2}^{+}$ & 3.990\\
& $\frac{1}{2}^{-}$  & 	3.097 & & & & $\frac{1}{2}^{+}$ & 4.005\\
& $\frac{3}{2}^{-}$  & 	3.087 & & & & $\frac{3}{2}^{+}$ & 4.000\\
& $\frac{5}{2}^{-}$  & 	3.074 & & & & $\frac{5}{2}^{+}$ & 3.993\\
&&&&&& $\frac{7}{2}^{+}$ & 	3.984\\
\noalign{\smallskip}
\cline{1-4}
\cline{6-9}
\noalign{\smallskip}
3P& $\frac{1}{2}^{-}$  &	3.361 & & & 1F & $\frac{5}{2}^{-}$ & 3.214\\
& $\frac{3}{2}^{-}$  & 	3.353 & & & & $\frac{7}{2}^{-}$ & 3.194\\
& $\frac{1}{2}^{-}$  & 	3.365 & & & & $\frac{3}{2}^{-}$ & 3.235\\
& $\frac{3}{2}^{-}$  & 	3.362 & & & & $\frac{5}{2}^{-}$ & 3.219\\
& $\frac{5}{2}^{-}$  & 	3.346 & & & & $\frac{7}{2}^{-}$ & 3.200\\
&&&&&& $\frac{9}{2}^{-}$ &  	3.177\\
\noalign{\smallskip}
\cline{1-4}
\cline{6-9}
\noalign{\smallskip}
4P& $\frac{1}{2}^{-}$ & 	3.608 & & & 2F & $\frac{5}{2}^{-}$ & 3.471\\
& $\frac{3}{2}^{-}$  &	        3.602 & & & & $\frac{7}{2}^{-}$ & 3.456\\
& $\frac{1}{2}^{-}$  &	        3.610 & & & & $\frac{3}{2}^{-}$ & 3.487\\
& $\frac{3}{2}^{-}$  & 	3.605 & & & & $\frac{5}{2}^{-}$ & 3.475\\
& $\frac{5}{2}^{-}$  & 	3.598 & & & & $\frac{7}{2}^{-}$ & 3.461\\
&&&&& & $\frac{9}{2}^{-}$ & 3.443\\
\noalign{\smallskip}
\cline{1-4}
\cline{6-9}
\noalign{\smallskip}
5P& $\frac{1}{2}^{-}$  &	3.841 & & & 3F & $\frac{5}{2}^{-}$ & 3.712\\
& $\frac{3}{2}^{-}$  & 	3.835 & & & & $\frac{7}{2}^{-}$ & 3.699\\
& $\frac{1}{2}^{-}$  & 	3.844 & & & & $\frac{3}{2}^{-}$ & 3.726\\
& $\frac{3}{2}^{-}$  & 	3.838 & & & & $\frac{5}{2}^{-}$ & 3.716\\
& $\frac{5}{2}^{-}$  & 	3.830 & & & & $\frac{7}{2}^{-}$ & 3.703\\
&&&&&& $\frac{9}{2}^{-}$ & 3.688\\
\noalign{\smallskip}
\cline{1-4}
\cline{6-9}
\noalign{\smallskip}
1D& $\frac{3}{2}^{+}$ &  	3.013 & & & 4F & $\frac{5}{2}^{-}$ & 3.937\\
& $\frac{5}{2}^{+}$ & 	2.997 & & & & $\frac{7}{2}^{-}$ & 3.926\\
& $\frac{1}{2}^{+}$ & 	3.031 & & & & $\frac{3}{2}^{-}$ & 3.949\\
& $\frac{3}{2}^{+}$ & 	3.019 & & & & $\frac{5}{2}^{-}$ & 3.940\\
& $\frac{5}{2}^{+}$ & 	3.003 & & & & $\frac{7}{2}^{-}$ & 3.930\\
& $\frac{7}{2}^{+}$ & 	2.983 & & & & $\frac{9}{2}^{-}$ & 3.917\\
\noalign{\smallskip}
\cline{1-4}
\cline{6-9}
\noalign{\smallskip}
2D& $\frac{3}{2}^{+}$ & 	3.288 & & & 5F & $\frac{5}{2}^{-}$ & 4.148\\
& $\frac{5}{2}^{+}$ & 	3.275 & & & & $\frac{7}{2}^{-}$ & 4.139 \\
& $\frac{1}{2}^{+}$ & 	3.302 & & & & $\frac{3}{2}^{-}$ & 4.157\\
& $\frac{3}{2}^{+}$ & 	3.292  & & & & $\frac{5}{2}^{-}$ & 4.150\\
& $\frac{5}{2}^{+}$ & 	3.280 & & & & $\frac{7}{2}^{-}$ & 4.142\\
& $\frac{7}{2}^{+}$ & 	3.264 & & & & $\frac{9}{2}^{-}$ & 4.132\\
\noalign{\smallskip}
\cline{1-4}
\cline{6-9}
\noalign{\smallskip}
3D& $\frac{3}{2}^{+}$ & 	3.540\\
& $\frac{5}{2}^{+}$ & 	3.530\\
& $\frac{1}{2}^{+}$ & 	3.551\\
& $\frac{3}{2}^{+}$ & 	3.544\\
& $\frac{5}{2}^{+}$ & 	3.534\\
& $\frac{7}{2}^{+}$ & 	3.522\\
\noalign{\smallskip}\hline
\end{tabular}
% Or use
 % with the correct table height
\end{table}

\begin{table}
\caption{Predicted masses of the orbital excited states of $\Sigma{_{c}^+}$ baryon (in GeV).}
%\scalebox{1.0}{
\label{tab:5}       % Give a unique label
\begin{tabular}{cccccccccccccc}
\hline\noalign{\smallskip}
State& $J^P$ & Present & \cite{Capstick1986}  & \cite{Ebert2008} & \cite{Ebert2011} & \cite{Yoshida2015} & \cite{Yamaguchi2015} & \cite{BChen2017} & PDG \cite{Tanabashi2018}\\
\noalign{\smallskip}\hline\noalign{\smallskip}
1P & $\frac{1}{2}^{-}$ &	2.806	  & 2.765 & 2.795 & 2.713 & 2.802 & 2.848 & 2.702 & 2.792${_{-0.005}^{+0.014}}$\\
& $\frac{3}{2}^{-}$  & 	2.794	  & 2.770 & 2.761 & 2.773 & 2.807 & 2.860 & 2.785 \\
& $\frac{1}{2}^{-}$  & 	2.812	  & 2.770 & 2.805 & 2.799 & & & 2.765\\
& $\frac{3}{2}^{-}$  & 	2.800	  & 2.805 & 2.798 & 2.798 & & 2.763 & 2.798 \\
& $\frac{5}{2}^{-}$  & 	2.783	  & 2.815 & 2.799 & 2.789 & 2.839 & 2.790 & 2.790  \\
\noalign{\smallskip}\hline\noalign{\smallskip}
2P& $\frac{1}{2}^{-}$ &	3.099	  & 3.185 & 3.176 & 3.125 & 2.826 & & 2.971 \\
& $\frac{3}{2}^{-}$  & 	3.089	  & 3.195 & 3.147 & 3.151 & 2.837 & & 3.036 \\
& $\frac{1}{2}^{-}$  & 	3.104	  & 3.195 & 3.186 & 3.172 & 3.018 & & 3.018\\
& $\frac{3}{2}^{-}$  & 	3.094	  & 3.210 & 3.180 & 3.172 & 3.044 & & 3.044\\
& $\frac{5}{2}^{-}$  & 	3.081	  & 3.220 & 3.167 & 3.161 & 3.316 & & 3.040\\
\noalign{\smallskip}\hline\noalign{\smallskip}
3P& $\frac{1}{2}^{-}$  &	3.366	  & & & 3.488 & 2.909\\
& $\frac{3}{2}^{-}$  & 	3.358	  & & & 3.486 & 2.910\\
& $\frac{1}{2}^{-}$  & 	3.370	  & & & 3.455\\
& $\frac{3}{2}^{-}$  & 	3.362	  & & & 3.469\\
& $\frac{5}{2}^{-}$  & 	3.352	  & & & 3.475 & 3.521\\
\noalign{\smallskip}\hline\noalign{\smallskip}
4P& $\frac{1}{2}^{-}$ & 	3.614	  & & & 3.770\\
& $\frac{3}{2}^{-}$  &	        3.608	  & & & 3.768\\
& $\frac{1}{2}^{-}$  &	        3.618	  & & & 3.743\\
& $\frac{3}{2}^{-}$  & 	3.611	  & & & 3.753\\
& $\frac{5}{2}^{-}$  & 	3.602	  & & & 3.757\\
\noalign{\smallskip}\hline\noalign{\smallskip}
5P& $\frac{1}{2}^{-}$  &	3.845	&		& \\
& $\frac{3}{2}^{-}$  & 	3.839	&		& \\
& $\frac{1}{2}^{-}$  & 	3.848	&		& \\
& $\frac{3}{2}^{-}$  & 	3.842	&		& \\
& $\frac{5}{2}^{-}$  & 	3.834	&		& \\
\noalign{\smallskip}\hline\noalign{\smallskip}
1D& $\frac{3}{2}^{+}$ &  	3.019	  & 3.060 & 3.005 & 3.043 & & 3.095 & 2.952 \\
& $\frac{5}{2}^{+}$ & 	3.004	  & 3.065 & 2.965 & 3.038 & 3.099 & 3.108 & 2.942\\
& $\frac{1}{2}^{+}$ & 	3.036	  & 3.005 & 3.014 & 3.041 & & 3.062 & 2.949 \\
& $\frac{3}{2}^{+}$ & 	3.025	  & 3.065 & 3.010 & 3.040 & & & 2.964\\
& $\frac{5}{2}^{+}$ & 	3.010	  & 3.080 & 3.001 & 3.023 & & 3.003 & 2.962 \\
& $\frac{7}{2}^{+}$ & 	2.999	  & 3.090 & 3.015 & 3.013 & & 3.015 & 2.943 \\
\noalign{\smallskip}\hline\noalign{\smallskip}
2D& $\frac{3}{2}^{+}$ & 	3.294	  & & & 3.366\\
& $\frac{5}{2}^{+}$ & 	3.281	  & & & 3.365 & 3.114\\
& $\frac{1}{2}^{+}$ & 	3.308	  & & & 3.370\\
& $\frac{3}{2}^{+}$ & 	3.299	  & & & 3.364\\
& $\frac{5}{2}^{+}$ & 	3.286	  & & & 3.349\\
& $\frac{7}{2}^{+}$ & 	3.270	  & & & 3.342\\
\noalign{\smallskip}\hline\noalign{\smallskip}
3D& $\frac{3}{2}^{+}$ & 	3.544	&		&\\
& $\frac{5}{2}^{+}$ & 	3.536	&		&\\
& $\frac{1}{2}^{+}$ & 	3.555	&		&\\
& $\frac{3}{2}^{+}$ & 	3.548	&		&\\
& $\frac{5}{2}^{+}$ & 	3.539	&		&\\
& $\frac{7}{2}^{+}$ & 	3.528	&		&\\
\noalign{\smallskip}\hline
\end{tabular}
% Or use
%\vspace*{5cm}  % with the correct table height
\end{table}

\begin{table}
\addtocounter{table}{-1}
\caption{to be continued...}
%\scalebox{1}{
\label{tab:5}       % Give a unique label
\begin{tabular}{ccccccccccc}
\hline\noalign{\smallskip}
State& $J^P$ & Present & \cite{Capstick1986}  & \cite{Ebert2008} & \cite{Ebert2011} & \cite{Yoshida2015} & \cite{Yamaguchi2015} & \cite{BChen2017} & PDG \cite{Tanabashi2018} \\
\noalign{\smallskip}\hline\noalign{\smallskip}
4D& $\frac{3}{2}^{+}$ & 	3.781	&		&\\
& $\frac{5}{2}^{+}$ & 	3.772	&		&\\
& $\frac{1}{2}^{+}$ & 	3.790	&		&\\
& $\frac{3}{2}^{+}$ & 	3.784	&		&\\
& $\frac{5}{2}^{+}$ & 	3.776	&		&\\
& $\frac{7}{2}^{+}$ & 	3.765	&		&\\
\noalign{\smallskip}\hline\noalign{\smallskip}
5D& $\frac{3}{2}^{+}$ & 	4.001	&		&\\
& $\frac{5}{2}^{+}$ & 	3.994	&		&\\
& $\frac{1}{2}^{+}$ & 	4.009	&		&\\
& $\frac{3}{2}^{+}$ & 	4.004	&		&\\
& $\frac{5}{2}^{+}$ & 	3.997	&		&\\
& $\frac{7}{2}^{+}$ & 	3.988	&		&\\
\noalign{\smallskip}\hline\noalign{\smallskip}
1F& $\frac{5}{2}^{-}$ &  	3.219	&&&	 3.283\\
& $\frac{7}{2}^{-}$ &  	3.200	&&&	 3.227\\
& $\frac{3}{2}^{-}$ &  	3.239	&&&	 3.288\\
& $\frac{5}{2}^{-}$ &  	3.225	&&&	 3.254\\
& $\frac{7}{2}^{-}$ &  	3.206	&&&	 3.253\\
& $\frac{9}{2}^{-}$ &  	3.184	&&&	 3.209\\
\noalign{\smallskip}\hline\noalign{\smallskip}
2F& $\frac{5}{2}^{-}$ &  	3.476	&\\
& $\frac{7}{2}^{-}$ &  	3.462	&\\
& $\frac{3}{2}^{-}$ &  	3.492	&\\
& $\frac{5}{2}^{-}$ &  	3.480	&\\
& $\frac{7}{2}^{-}$ &  	3.466	&\\\
& $\frac{9}{2}^{-}$ &  	3.449	&\\
\noalign{\smallskip}\hline\noalign{\smallskip}
3F& $\frac{5}{2}^{-}$ &  	3.716	&\\
& $\frac{7}{2}^{-}$ &  	3.704	&\\
& $\frac{3}{2}^{-}$ &  	3.730	&\\
& $\frac{5}{2}^{-}$ &  	3.720	&\\
& $\frac{7}{2}^{-}$ &  	3.708	&\\
& $\frac{9}{2}^{-}$ &  	3.693	&\\
\noalign{\smallskip}\hline\noalign{\smallskip}
4F& $\frac{5}{2}^{-}$ &  	3.941	&\\
& $\frac{7}{2}^{-}$ &  	3.930	&\\
& $\frac{3}{2}^{-}$ &  	3.952	&\\
& $\frac{5}{2}^{-}$ &  	3.944	&\\
& $\frac{7}{2}^{-}$ &  	3.934	&\\
& $\frac{9}{2}^{-}$ &  	3.921	&\\
\noalign{\smallskip}\hline\noalign{\smallskip}
5F& $\frac{5}{2}^{-}$ &	4.152	&\\
& $\frac{7}{2}^{-}$ &  	4.143	&\\
& $\frac{3}{2}^{-}$ &  	4.161	&\\
& $\frac{5}{2}^{-}$ &  	4.154	&\\
& $\frac{7}{2}^{-}$ &  	4.146	&\\
& $\frac{9}{2}^{-}$ &  	4.135	&\\
\noalign{\smallskip}\hline
\end{tabular}
% Or use
%\vspace*{5cm}  % with the correct table height
\end{table}

\begin{table}
\caption{Predicted masses of the orbital excited states of $\Sigma{_{c}^0}$ baryon (in GeV).}
%\scalebox{0.75}{
\label{tab:6}       % Give a unique label
\begin{tabular}{ccccccccccc}
\hline\noalign{\smallskip}
State& $J^P$ & Present & PDG \cite{Tanabashi2018} & & State& $J^P$ & Present & PDG \cite{Tanabashi2018} \\
\noalign{\smallskip}
\cline{1-4}
\cline{6-9}
\noalign{\smallskip}
1P & $\frac{1}{2}^{-}$ &	2.812	&	 2.806${_{-0.007}^{+0.005}}$ & & 4D & $\frac{3}{2}^{+}$ & 3.786 &\\
& $\frac{3}{2}^{-}$  & 	2.801	&	& & & $\frac{5}{2}^{+}$ & 3.777 &\\
& $\frac{1}{2}^{-}$  & 	2.818	&	& & & $\frac{1}{2}^{+}$ & 3.796 &\\
& $\frac{3}{2}^{-}$  & 	2.807	&	& & & $\frac{3}{2}^{+}$ & 3.789 &\\
& $\frac{5}{2}^{-}$  & 	2.792	&	& & & $\frac{5}{2}^{+}$ & 3.781	&\\
&&&&&& $\frac{7}{2}^{+}$ & 3.770 &\\
\noalign{\smallskip}
\cline{1-4}
\cline{6-9}
\noalign{\smallskip}
2P& $\frac{1}{2}^{-}$ &	3.106	& & & 5D& $\frac{3}{2}^{+}$ & 4.006 &\\
& $\frac{3}{2}^{-}$  & 	3.096	& & & & $\frac{5}{2}^{+}$ & 3.998 &\\
& $\frac{1}{2}^{-}$  & 	3.110	& & & & $\frac{1}{2}^{+}$ & 4.014 &\\
& $\frac{3}{2}^{-}$  & 	3.101	& & & & $\frac{3}{2}^{+}$ & 4.008 &\\
& $\frac{5}{2}^{-}$  & 	3.088	& & & & $\frac{5}{2}^{+}$ & 4.001 &\\
&&&&&& $\frac{7}{2}^{+}$ &      3.992 &\\
\noalign{\smallskip}
\cline{1-4}
\cline{6-9}
\noalign{\smallskip}
3P& $\frac{1}{2}^{-}$  &	3.373	& & & 1F& $\frac{5}{2}^{-}$ & 3.224\\
& $\frac{3}{2}^{-}$  & 	3.365	& & & & $\frac{7}{2}^{-}$ & 3.206\\
& $\frac{1}{2}^{-}$  & 	3.377	& & & & $\frac{3}{2}^{-}$ & 3.244\\
& $\frac{3}{2}^{-}$  & 	3.369	& & & & $\frac{5}{2}^{-}$ & 3.230\\
& $\frac{5}{2}^{-}$  & 	3.358	& & & & $\frac{7}{2}^{-}$ & 3.212\\
&&&&&& $\frac{9}{2}^{-}$ &  	3.190 &\\
\noalign{\smallskip}
\cline{1-4}
\cline{6-9}
\noalign{\smallskip}
4P& $\frac{1}{2}^{-}$ & 	3.620 	& & & 2F & $\frac{5}{2}^{-}$ & 3.482\\
& $\frac{3}{2}^{-}$  &	        3.613	& & & & $\frac{7}{2}^{-}$ & 3.467\\
& $\frac{1}{2}^{-}$  &	        3.623	& & & & $\frac{3}{2}^{-}$ & 3.499\\
& $\frac{3}{2}^{-}$  & 	3.616	& & & & $\frac{5}{2}^{-}$ & 3.487\\
& $\frac{5}{2}^{-}$  & 	3.607	& & & & $\frac{7}{2}^{-}$ & 3.477\\
&&&&&& $\frac{9}{2}^{-}$ &  	3.453 &\\
\noalign{\smallskip}
\cline{1-4}
\cline{6-9}
\noalign{\smallskip}
5P& $\frac{1}{2}^{-}$  &	3.850	& & & 3F & $\frac{5}{2}^{-}$ & 3.720\\
& $\frac{3}{2}^{-}$  & 	3.844	& & & & $\frac{7}{2}^{-}$ & 3.709\\
& $\frac{1}{2}^{-}$  & 	3.852	& & & & $\frac{3}{2}^{-}$ & 3.733\\
& $\frac{3}{2}^{-}$  & 	3.847	& & & & $\frac{5}{2}^{-}$ & 3.724\\
& $\frac{5}{2}^{-}$  & 	3.839	& & & & $\frac{7}{2}^{-}$ & 3.712\\
&&&&&& $\frac{9}{2}^{-}$ &  	3.698 &\\
\noalign{\smallskip}
\cline{1-4}
\cline{6-9}
\noalign{\smallskip}
1D& $\frac{3}{2}^{+}$ &  	3.026	& & & 4F & $\frac{5}{2}^{-}$ & 3.945\\
& $\frac{5}{2}^{+}$ & 	3.012	& & & & $\frac{7}{2}^{-}$ & 3.935\\
& $\frac{1}{2}^{+}$ & 	3.043	& & & & $\frac{3}{2}^{-}$ & 3.956\\
& $\frac{3}{2}^{+}$ & 	3.032	& & & & $\frac{5}{2}^{-}$ & 3.948\\
& $\frac{5}{2}^{+}$ & 	3.017	& & & & $\frac{7}{2}^{-}$ & 3.938\\
& $\frac{7}{2}^{+}$ & 	2.999	& & & & $\frac{9}{2}^{-}$ & 3.925\\
\noalign{\smallskip}
\cline{1-4}
\cline{6-9}
\noalign{\smallskip}
2D& $\frac{3}{2}^{+}$ & 	3.299	& & & 5F & $\frac{5}{2}^{-}$ & 4.155\\
& $\frac{5}{2}^{+}$ & 	3.287	& & & & $\frac{7}{2}^{-}$ & 4.146\\
& $\frac{1}{2}^{+}$ & 	3.313	& & & & $\frac{3}{2}^{-}$ & 4.165\\
& $\frac{3}{2}^{+}$ & 	3.304	& & & & $\frac{5}{2}^{-}$ & 4.158\\
& $\frac{5}{2}^{+}$ & 	3.292	& & & & $\frac{7}{2}^{-}$ & 4.149\\
& $\frac{7}{2}^{+}$ & 	3.276	& & & & $\frac{9}{2}^{-}$ & 4.139\\
\noalign{\smallskip}
\cline{1-4}
\cline{6-9}
\noalign{\smallskip}
3D& $\frac{3}{2}^{+}$ & 	3.551	&\\
& $\frac{5}{2}^{+}$ & 	3.541	&\\
& $\frac{1}{2}^{+}$ & 	3.563	&\\
& $\frac{3}{2}^{+}$ & 	3.555	&\\
& $\frac{5}{2}^{+}$ & 	3.545	&\\
& $\frac{7}{2}^{+}$ & 	3.532	&\\
\noalign{\smallskip}\hline
\end{tabular}
% Or use
%\vspace*{5cm}  % with the correct table height
\end{table}

\noindent The first order correction $ V_I(x) $ can be written in the form of Casimir charges of the fundamental and the adjoint representation such as $ C_F = \frac{2}{3} $ and $ C_A = 3 $ respectively,

\begin{equation}
\label{eqn:10} 
V_I(x) = - C_F C_A \frac{\alpha{_{S}^2}}{4x^2}.
\end{equation}

The spin-dependent potential $ V_{SD}(x) $ determine the mass difference between degenerate baryonic states given by \cite{Voloshin2008,Bijker1994,Bijker1998},

\begin{equation}
\label{eqn:11} 
V_{SD}(x) = V_{SS}(x)(\vec{S_\rho}\cdot \vec{S_\lambda}) + V_{\gamma S}(x)(\vec{L} \cdot \vec{S}) + V_{T}(x) \left[S^2 - \frac{3(\vec S \cdot \vec x)(\vec S \cdot \vec x)}{x^2}\right],
\end{equation}

where $ V_{SS} (x) $, $ V_{\gamma S} (x) $ and $ V_{T} (x) $ are the spin-spin, spin-orbit interaction and the spin-tensor interaction terms (for details see Ref. \cite{Shah2016epja}) .\\ 

\noindent The six-dimensional hyperradial Schr\"{o}dinger equation is solved numerically using Mathematica notebook \cite{Lucha1999},

\begin{equation}
\label{eqn:18}  
\left[\frac{1}{2m}\left(-\frac{d^2}{dr^2}+\frac{\frac{15}{4}+\gamma(\gamma+4)}{r^2}\right)+V(x)\right]\phi_{\gamma}(x) = E_B \phi_{\gamma}(x).
\end{equation}

\noindent Here, m is the reduced mass (see in Eq. (\ref{eqn:1})) and $ E_B $ gives the binding energy of the baryonic states, and $V(x)=V_{SI}(x)+V_{SS}(x)+V_{L S}(x)+V_{T}(x)$, is the total potential of the baryonic system.\\

The spin average masses are determined by taking a summation of model quark masses with its binding energy,

\begin{equation}
\label{eqn:19}
M_{SA} = E_B + m_{q_1} + m_{q_2} + m_{q_3}.
\end{equation}

\noindent Therefore, the total mass is,

\begin{equation}
\label{eqn:20}  
M_{total} = M_{SA} + V(x) - V_{SI}(x).
\end{equation}

\noindent In this way, we calculate the excited state masses of the nonstrange singly charmed baryons. We analyze them in the next section. 

\label{sec:3}

\begin{figure*}
\resizebox{1.0\textwidth}{!}{%
  \includegraphics{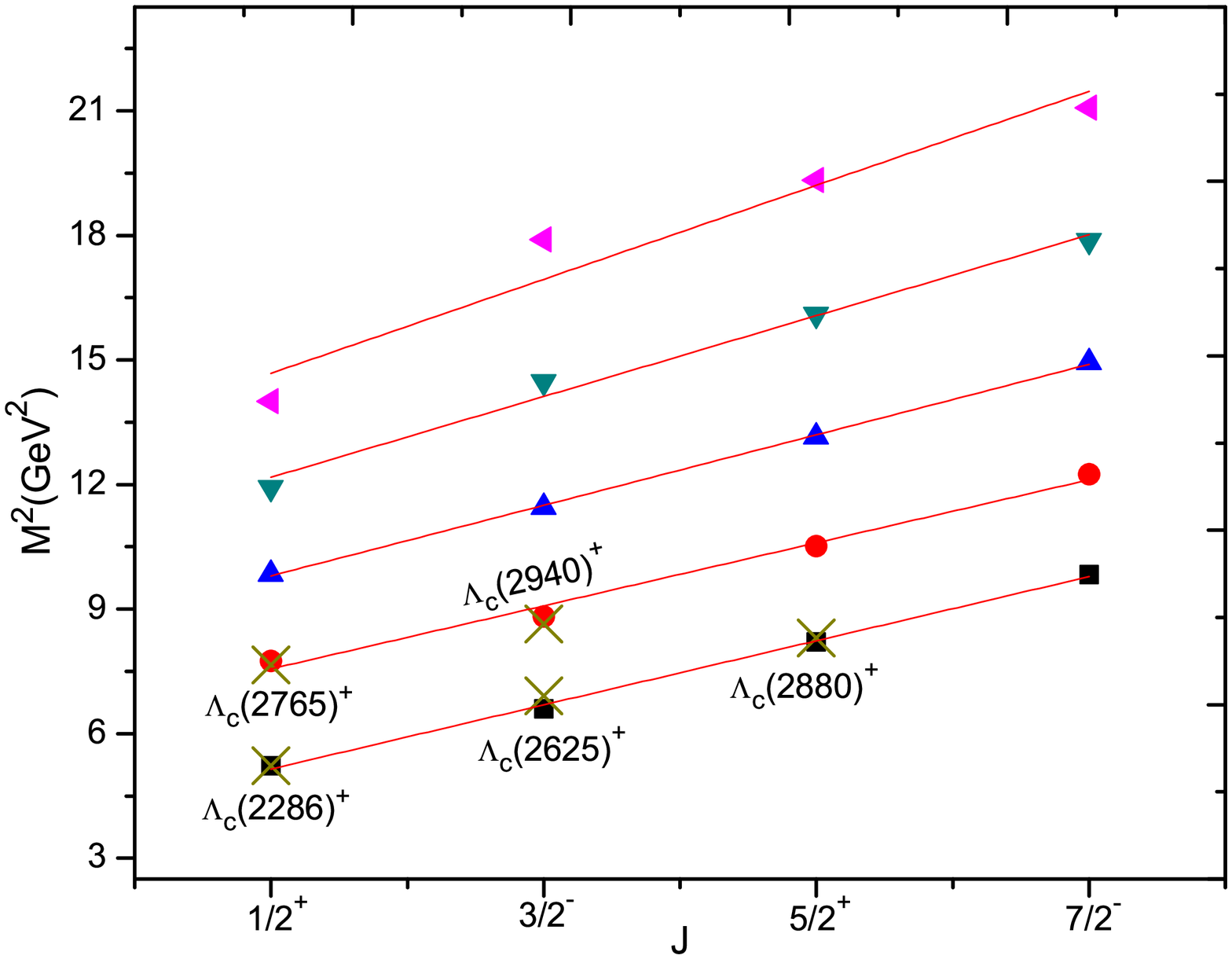}
}
\caption{The $(M^2 \rightarrow J)$ Regge trajectory of $\Lambda{_{c}^+}$ baryon with natural parity. An experimental available measurements are presented by cross sign.}
\label{Fig:2}       % Give a unique label
\end{figure*}

\begin{figure*}
\resizebox{1.0\textwidth}{!}{%
  \includegraphics{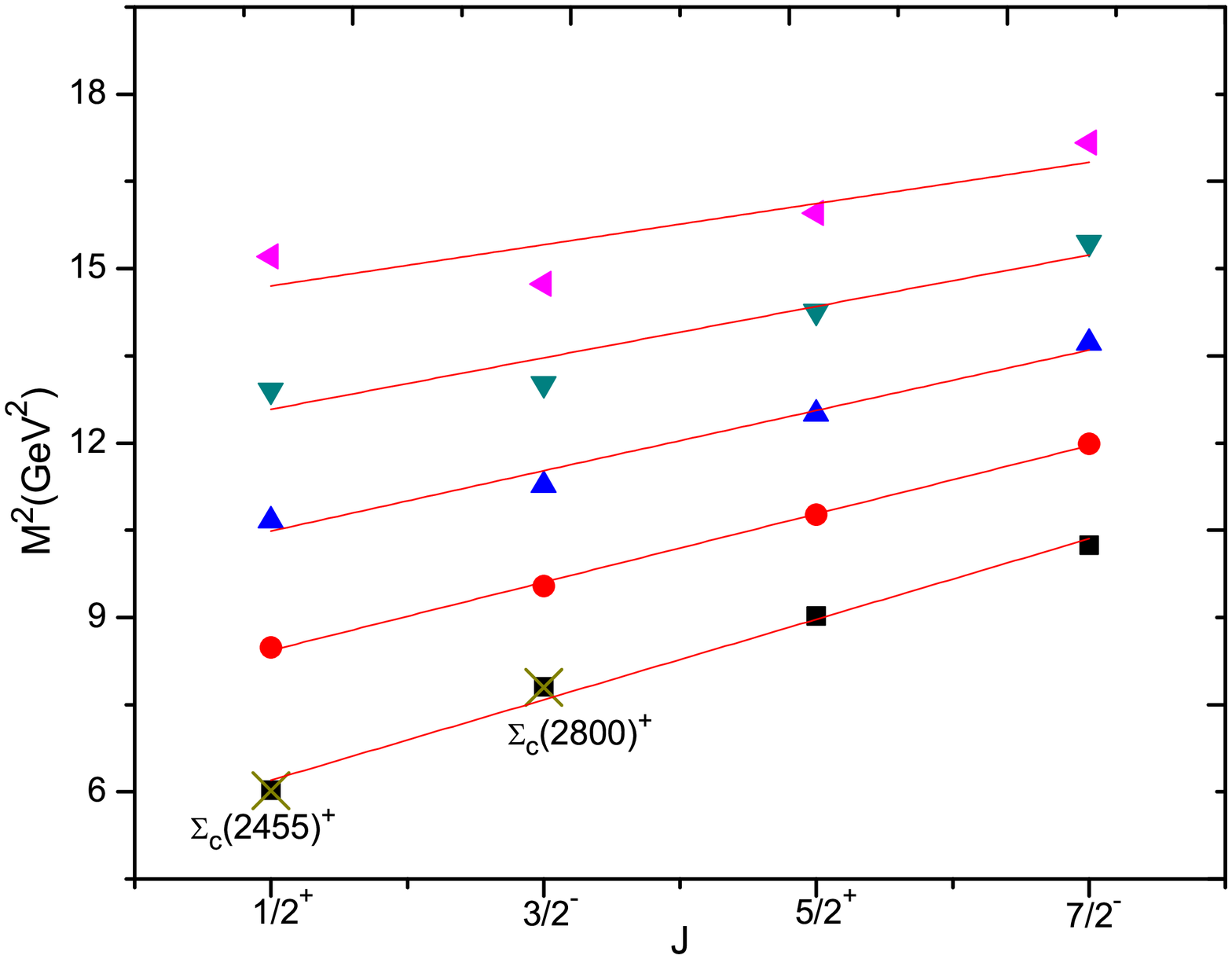}
   \includegraphics{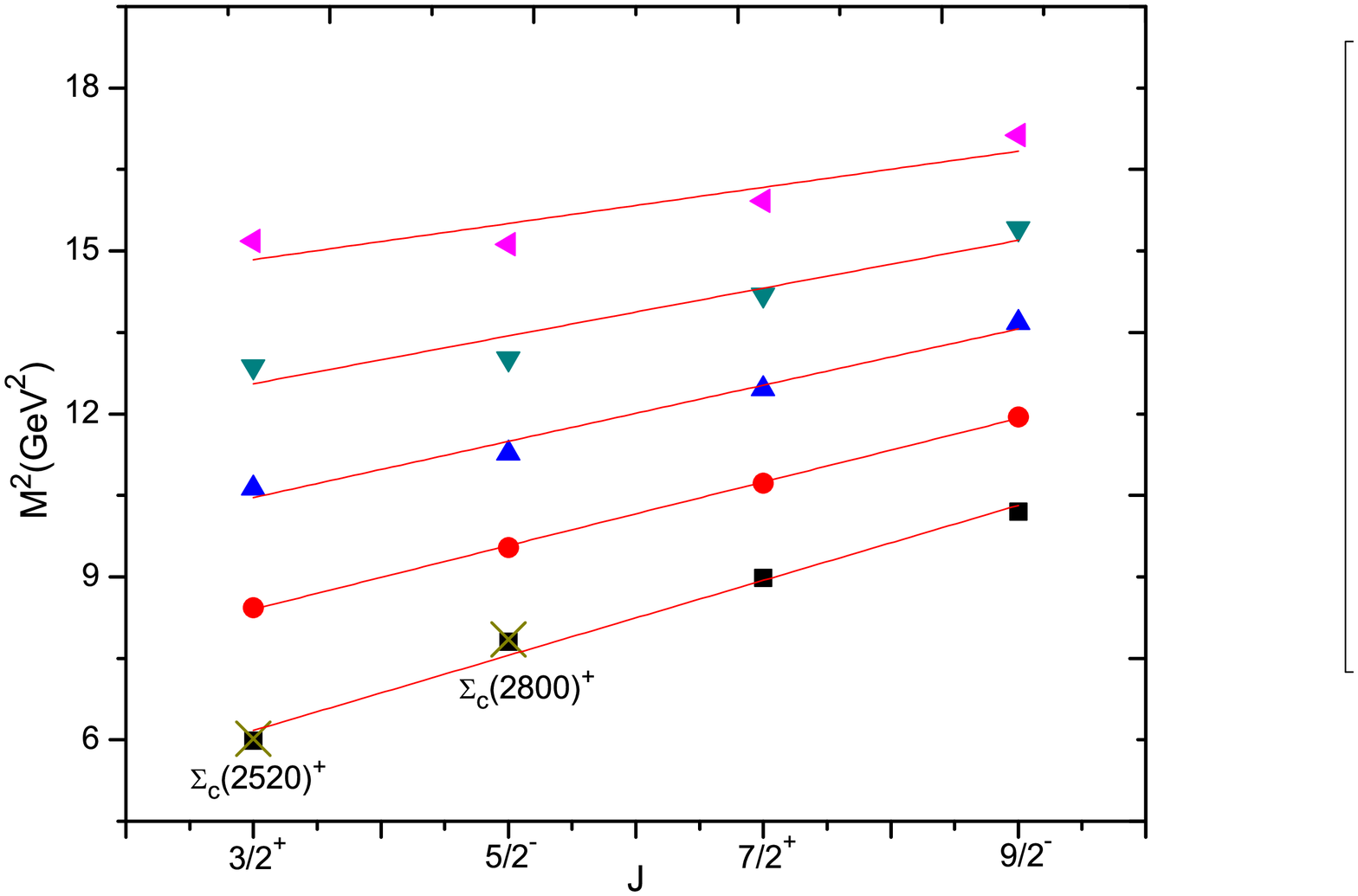}
}
\caption{The $(M^2 \rightarrow J)$ Regge trajectory of $\Sigma{_{c}^+}$ baryon with natural parity (left) and unnatural parity (right).}
\label{Fig:3}       % Give a unique label
\end{figure*}

\section{Mass Spectra and Regge Trajectories}

The masses of the radial and the orbital excited states of nonstrange singly charmed baryons are calculated in the non-relativistic framework of hypercentral Constituent Quark Model (hCQM). Here, the screened potential is used as a confining potential with first order correction. Our calculated masses are presented in Tables \ref{tab:2}-\ref{tab:6} with other theoretical predictions and known experimental observations. Our results are used to draw Regge trajectories in the $(J, M^2)$ plane (see in Figs. \ref{Fig:2} and \ref{Fig:3}). Regge trajectories can be used to determine the possible quantum number of the particular hadronic state \cite{Guo2008}.\\

The Regge trajectories are plotted in $(J, M^2)$ plane with natural and unnatural parities given by $P= (-1)^{J-\frac{1}{2}}$ and $P= (-1)^{J+\frac{1}{2}}$ respectively. Hence, the states $J^P={\frac{1}{2}}^+$, ${\frac{3}{2}}^-$, ${\frac{5}{2}}^+$ and ${\frac{7}{2}}^-$ are available with the natural parities and the states $J^P={\frac{3}{2}}^+$, ${\frac{5}{2}}^-$, ${\frac{7}{2}}^+$ and ${\frac{9}{2}}^-$ are for the unnatural parities. Using the equation, 
 
\begin{equation}
\label{eqn:21}  
J=\alpha M^2+\alpha_0,
\end{equation}

\noindent we construct the Regge trajectories in the $(J, M^2)$ plane. Here, $\alpha$ and $\alpha_0$ are the fitted slopes and the intercepts respectively. The straight lines are obtained by the linear fitting. For the indication our calculated masses are presented by various mathematical symbols and the experimental observations by cross sign with a particle name.

\subsection{${\Lambda{_{c}^+}}$ states}

We fix the ground state $(1S)$ mass of $\Lambda{_{c}^+}$ baryon with the experimentally known value 2.286 GeV \cite{Tanabashi2018} and then calculate its excited state mass spectra. For $2S$ state with $J^P = {\frac{1}{2}}^+$, our result is 2.785 GeV, close to PDG \cite{Tanabashi2018} by a mass difference of 19 MeV and also in agreement with the theoretical predictions \cite{Ebert2008,Ebert2011,Chen2015,Yoshida2015,Yamaguchi2015,BChen2017}. For the higher radial excitation ($3S$, $4S$ and $5S$ states), our results are in accordance with the predictions of D. Ebert \textit{et al}. \cite{Ebert2011} (see in Table \ref {tab:2}).\\ 

In 2011, CDF \cite{Aaltonen2011} measured the masses, 2592.25 $\pm$ 0.24(stat) $\pm$ 0.14(syst) MeV$/c^2$ and 2628.11 $\pm$ 0.13(stat) $\pm$ 0.14(syst) MeV$/c^2$, from their decay into ${\Lambda{_{c}^+}} \pi^+ \pi^-$. These states are assigned with the quantum numbers $J^P = {\frac{1}{2}}^-$ and ${\frac{3}{2}}^-$ respectively. For the $1P$ state with $J^P = {\frac{1}{2}}^-$ and ${\frac{3}{2}}^-$, our predictions are 2.573 GeV and 2.568 GeV respectively, which are smaller than Ref. \cite{Tanabashi2018} and the theoretical predictions \cite{Ebert2008,Ebert2011,Chen2015,Yoshida2015,Yamaguchi2015,BChen2017}. Here, 2.573 GeV with $J^P = {\frac{1}{2}}^-$ is in good agreement with the lattice QCD result 2.578 GeV \cite{Rubio2015}. For $2P$ state with $J^P = {\frac{1}{2}}^-$ and ${\frac{3}{2}}^-$, our predicted masses are 2.978 GeV and 2.970 GeV respectively, which are close to the results of Refs. \cite{Ebert2008,Ebert2011,Chen2015,BChen2017,Shah2016epja} as well as the result of ${\frac{3}{2}}^-$ is reasonably close to the experimental observation 2939.6 MeV \cite{Tanabashi2018}. Our predictions of $3P$, $4P$ and $5P$ states are larger than Ref. \cite{Ebert2011}.\\     
     
In 2007, the \textit{BABAR} Collaboration \cite{Aubert2007} measured a state with mass 2881.9 $\pm$ 0.1(stat) $\pm$ 0.5(syst) MeV$/c^2$ and then in the same year Belle Collaboration \cite{Mizuk2007} identified its spin-parity as $J^P = {\frac{5}{2}}^+$. At latest, the LHCb Collaboration \cite{Aaij2017} observed the mass and the quantum number of $ {\Lambda{_{c}(2880)^+}} $ baryon as: 2881.75 $\pm$ 0.29(stat) $\pm$ 0.07(syst) MeV, $J^P={\frac{5}{2}}^+$. They also found a new resonance state $ {\Lambda{_{c}(2860)^+}} $ having a mass 2856.1${^{+2.0}_{-1.7}}$(stat) $\pm$ 0.5(syst) MeV with $J^P={\frac{3}{2}}^+$. For these states $ {\Lambda{_{c}(2880)^+}} $ and $ {\Lambda{_{c}(2860)^+}} $, our results are 2.876 GeV and 2.865 GeV having mass difference of 20 MeV and 16 MeV with Ref. \cite{Tanabashi2018}. For the $2D$-state, our predictions are 3.256 GeV and 3.244 GeV for spin-parity ${\frac{3}{2}}^+$ and ${\frac{5}{2}}^+$ respectively, which are in accordance with the results of Ref. \cite{Ebert2008} and larger than the other theoretical predictions \cite{Ebert2011,Yoshida2015,Shah2016epja,Yamaguchi2015}. Still there are no experimental evidences available for the $F$-states. Our predictions of the $1F$ states are 3.152 GeV and 3.136 GeV for the spin-parity ${\frac{5}{2}}^-$ and ${\frac{7}{2}}^-$ respectively, which are compatible with predictions of Refs. \cite{Ebert2008,Ebert2011,Chen2015,Yamaguchi2015,BChen2017}. For $3D$, $4D$, $2F$ and $4F$ states, our results are overestimated from Refs. \cite{Ebert2011,Yoshida2015}. By taking a comparison with Ref. \cite{Ebert2011} the mass differences are increasing with $n$ (principle quantum number).\\

Fig. (\ref{Fig:2}) shows that the Regge trajectories of the $ \Lambda{_{c}^+} $ baryon and they are almost parallel and equidistant. The square masses of the $ \Lambda{_{c}^+} $ baryon fit nicely to the linear trajectories in $(J, M^2)$ plane. The available experimental data are well matched with the corresponding Regge trajectories obtained in our model. Still the quantum state of $ {\Lambda{_{c}(2765)^+}} $ baryon is not confirmed experimentally. On the Regge line such a state $ {\Lambda{_{c}(2765)^+}} $ is following $ {\Lambda{_{c}(2940)^+}} $ state in $(J, M^2)$ plane. So it may have $2S$ state with spin $\frac{1}{2}$.

\subsection{$\Sigma{_{c}}$ states}

Just after the discovery of $\Lambda{_{c}^+}$ baryon at Brookhaven National Laboratory (BNL) in 1975 \cite{Cazzoli1975}, the Fermi National Accelerator Laboratory (FNAL) found the evidence of $\Sigma{_{c}}$ baryons \cite{Knapp1976}. It has three isospin states with different quarks constitutions: $\Sigma{_{c}^{++}}$ $(uuc)$, $\Sigma{_{c}^{+}}$ $(udc)$ and $\Sigma{_{c}^{0}}$ $(ddc)$. In the present study, their masses are calculated separately by considering unequal light quarks $(u$ and $d)$ masses. The Refs. \cite{Capstick1986,Ebert2008,Yoshida2015,Yamaguchi2015,BChen2017} consider the same light quarks masses, so here our predictions are compared only with $\Sigma{_{c}^{+}}$ states.\\  

We fix the ground state $(1S)$ of isotriplet $\Sigma{_{c}}$ baryons with experimental value from Ref. \cite{Tanabashi2018}, and calculate their respective radial and orbital excited states. From the quark mass hierarchy, the down ($d$) quark is heavier than up ($u$) quark, so the masses of the $\Sigma{_{c}^{0}}$ baryons are expected to be higher than the masses of  $\Sigma{_{c}^{++}}$ baryons. Many experimental observations \cite{Aaltonen2011,Artuso2002,Athar2005,Lee2014} gives negative mass splittings of $ M(\Sigma{_{c}(2455)^{0}} - \Sigma{_{c}(2455)^{++}})$ and $ M(\Sigma{_{c}(2520)^{0}} - \Sigma{_{c}(2520)^{++}})$. Expected by the models (for details see Ref. \cite{Silvestre-Brac2003}), in our case both these isospin mass splittings are positive. $\Sigma{_{c}^{+}}$ has been observed in a decay mode $\Lambda_c^+ \pi^0$ by single experiment \cite{Calicchio1980}, because the detectors are inefficient for the $\pi^0$ identification as a decay product. The isotriplet $\Sigma{_{c}}$ baryons have two spin states: $\frac{1}{2}$ and $\frac{3}{2}$. For the $2S$ state with $J^P = {\frac{1}{2}}^+$ and $J^P = {\frac{3}{2}}^+$, our results are 2.912 GeV and 2.951 GeV, consistent with the results 2.901 GeV and 2.936 GeV of D. Ebert \textit{et al.} \cite{Ebert2011}. For the $3S$ state, the mass corresponding to $J^P = {\frac{1}{2}}^+$ is 3.266 GeV, which is nearer to 3.271 GeV of \cite{Ebert2011}, and for $J^P = {\frac{3}{2}}^+$ it is 3.288 GeV, overestimated from Refs. \cite{Capstick1986,Ebert2008,Yoshida2015,Yamaguchi2015,BChen2017} (see in Table \ref {tab:3}).\\

Experimentally, only the first orbital excited states of isotriplet $\Sigma{_{c}}$ baryons are observed \cite{Athar2005,Mizuk2005} and still their $J^P$ values are not known. Our calculated orbital excited state masses of the isotriplet $\Sigma{_{c}}$ baryons are presented in Tables \ref {tab:4}-\ref {tab:6}.  For the $1P$, $2P$ and $1D$ states of $\Sigma{_{c}^{+}}$ baryon our predictions are in agreement with the results of Refs. \cite{Capstick1986,Ebert2008,Ebert2011,BChen2017} and, for the $3P$,$4P$, $2D$ and $1F$ states our results are smaller than \cite{Ebert2011} (see in Table \ref {tab:5}). That means, the screening effect comes into the picture that gives the lower mass of the higher excited states.\\

Fig. (\ref{Fig:3}) shows the Regge trajectories of $\Sigma{_{c}^{+}}$ baryon with natural and unnatural parities in the $(J, M^2)$ plane. For the massive states of $\Sigma{_{c}^{+}}$ baryon the trajectories are to become horizontal, which leads to smaller slopes for heavy quarks (for more study see Refs. \cite{Guo2008,Zhang2005,Li2004,Brisudova2000}). From the spectroscopy and the Regge trajectories here we are unable to predict the possible $J^P$ value of the isotriplet $\Sigma{_{c}}(2800)$ baryons.   

\section{Properties}
\label{sec:4}

\subsection{Strong Decays}

We are using our calculated masses for the determination of the decay properties, that is important to identify the $J^P$ value of newly observed experimental states. As stated in the introduction, singly charmed baryons have one charm quark and two light quarks. So it provides an excellent base for rectifying the heavy quark symmetry of the heavy quark and the chiral symmetry of the light quarks in the low energy regime. A Heavy Hadron Chiral Perturbation Theory (HHChPT) represents the chiral Lagrangian in which the heavy quark symmetry and the chiral symmetry are incorporated (see Refs. \cite{Yan1992,Cho1994,Cheng1997,Pirjol1997}). Such a Lagrangian describes the strong interactions of heavy baryons with the emission of light pseudoscalar mesons. It contains strong coupling constants: $g_1$ and $g_2$ for the $P$-wave transitions, $h_2$ to $h_7$ for the $S$-wave transitions and $h_8$ to $h_{15}$ used for the $D$-wave transitions. The expressions of partial decay widths are taken from Refs. \cite{Pirjol1997,Cheng2007,Cheng2015,Cheng20151,Gandhi2018} and the pion decay constant is, $f_{\pi}$ = 130.2 MeV \cite{Tanabashi2018}.\\ 

\subsubsection{$P$-wave transitions}

\begin{table}
\caption{The strong $P$-wave transitions (in MeV) of the $s$-wave baryons.}
\scalebox{1}{
\label{tab:7}       % Give a unique label
\begin{tabular}{cccccccccccccccc}
\hline\noalign{\smallskip}
Decay & HHChPT & HHChPT & \cite{Cheng20151} & \cite{Cheng2007} & \cite{Gandhi2018} & \cite{Kawakami2018} & \cite{HCKim2019} & \cite{Ivanov1999} & \cite{Tawfiq1998} & \cite{Albertus2005} & PDG\cite{Tanabashi2018} \\
& Present & PDG-2018 & &&&&&&&&\\
\noalign{\smallskip}\hline\noalign{\smallskip}
$\Sigma_c(2455)^{++} \rightarrow \Lambda{_{c}^+} \pi^+$ & 1.24 & input & input & input & 2.34 & 1.96 & 1.93 &  2.85 & 1.64 & 2.41 & 1.89\\
&${_{-0.12}^{+0.06}}$&&&&&${_{-0.14}^{+0.07}}$&& $\pm$ 0.19 && $\pm$ 0.07 & ${_{-0.18}^{+0.09}}$\\
\noalign{\smallskip}
$\Sigma_c(2455)^{+} \rightarrow \Lambda{_{c}^+} \pi^0$ & 2.62 & 2.20 & 2.3 & 2.6 & 2.59 & 2.28 & 2.24 & 3.63 & 1.70 & 2.79 & $<$4.6 \\
&${_{-0.25}^{+0.12}}$&${_{-0.24}^{+0.14}}$&${_{-0.2}^{+0.1}}$& $\pm$ 0.04 &&${_{-0.17}^{+0.09}}$&& $\pm$ 0.27 && $\pm$ 0.08\\
\noalign{\smallskip}
$\Sigma_c(2455)^{0} \rightarrow \Lambda{_{c}^+} \pi^-$ & 3.45 & 1.87 & 1.9 & 2.2 & 2.21 & 1.94 & 1.90 & 2.65 & 1.57 & 2.37 & 1.83 \\
&${_{-0.33}^{+0.16}}$&${_{-0.09}^{+0.18}}$&${_{-0.2}^{+0.1}}$& $\pm$ 0.3 &&${_{-0.14}^{+0.07}}$&& $\pm$ 0.19 && $\pm$ 0.07 & ${_{-0.19}^{+0.11}}$\\
\noalign{\smallskip}
$\Sigma{_{c}(2520)^{++}} \rightarrow \Lambda{_{c}^+} \pi^+$ & 12.61 & 14.20 & 14.5 & 16.7 & 21.34 & 14.7 & 14.7 & 21.99 & 12.84 & 17.52 & 14.78 \\
&${_{-1.20}^{+0.60}}$&${_{-1.36}^{+0.70}}$&${_{-0.8}^{+0.5}}$& $\pm$ 2.3 &&${_{-1.1}^{+0.06}}$&& $\pm$ 0.87 && $\pm$ 0.75 & ${_{-0.40}^{+0.30}}$\\
\noalign{\smallskip}
$\Sigma{_{c}(2520)^{+}} \rightarrow \Lambda{_{c}^+} \pi^0$ & 15.85 & 14.74 & 15.2 & 17.4 & 12.83 & 15.3 & 15.02 & & & 15.31 & $<$17\\
&${_{-1.51}^{+0.75}}$&${_{-1.94}^{+1.33}}$&${_{-1.3}^{+0.6}}$& $\pm$ 2.3 &&${_{-1.1}^{+0.6}}$&&&& $\pm$ 0.74\\
\noalign{\smallskip}
$\Sigma{_{c}(2520)^{0}} \rightarrow \Lambda{_{c}^+} \pi^-$ & 17.69 & 14.22 & 14.7 & 16.6 & 20.97 & 14.7 & 14.49 & 21.21 & 12.40 & 16.90 & 15.3 \\
&${_{-1.68}^{+0.84}}$&${_{-1.37}^{+0.69}}$&${_{-1.2}^{+0.6}}$& $\pm$ 2.2 &&${_{-1.1}^{+0.06}}$&& $\pm$ 0.81 && $\pm$ 0.72 & ${_{-0.5}^{+0.4}}$ \\
\noalign{\smallskip}\hline
\end{tabular}}
{\tiny{Ref. \cite{Cheng20151} $\rightarrow$ HHChPT(2015) PDG-2014 and\\
Ref. \cite{Cheng2007} $\rightarrow$ HHChPT(2007) PDG-2006.}}
% Or use
%\vspace*{5cm}  % with the correct table height
\end{table}

The $P$-wave couplings take place among the $s$-wave baryons. For the decay $\Sigma_c(2520) \rightarrow \Sigma_c(2455) \pi$, experimentally the mass difference between these two singly charmed baryons is around 65 MeV \cite{Tanabashi2018}. A single pion do not have such amount of phase space for this decay. Therefore, such a decay is kinematically prohibited and we cannot extract coupling constant $g_1$ here. The $g_2$ can be obtained from the allowed decay channels: $\Sigma_c(2455)^{++} \rightarrow \Lambda{_{c}^+} \pi^+$, $\Sigma_c(2455)^0 \rightarrow \Lambda{_{c}^+} \pi^-$, $\Sigma_c(2520)^{++} \rightarrow \Lambda{_{c}^+} \pi^+$ and $\Sigma_c(2520)^0 \rightarrow \Lambda{_{c}^+} \pi^-$ as     

\begin{equation}
\label{eqn:37} 
%\begin{aligned}
\mid g_2 \mid = 0.550{_{-0.027}^{+0.013}}, \hspace{0.4cm} 0.544{_{-0.018}^{+ 0.010}}, \hspace{0.4cm} 0.561{_{-0.007}^{+ 0.005}}, \hspace{0.4cm} 0.570{_{-0.009}^{+ 0.007}}
%\end{aligned}
\end{equation}

\noindent respectively, using their masses and respective decay widths from PDG-2018 \cite{Tanabashi2018}. Therefore, an average $\mid g_2 \mid_{2018}$ = $0.566{_{-0.027}^{+0.013}}$ (using PDG-2018 \cite{Tanabashi2018}), which is nearer to $\mid g_2 \mid_{2015}$ = $0.565{_{-0.024}^{+0.011}}$ of Ref. \cite{Cheng2015} and $\mid g_2 \mid_{2006}$ = $0.591 \pm 0.023$ of Ref. \cite{Cheng2007}. The non-relativistic quark model determined the coupling constants $g_1$ and $g_2$ in the form of an axial-vector coupling $g{_{A}^q}$ in a single light quark transition $u \rightarrow d$, written as \cite{Pirjol1997}

\begin{equation}
\label{eqn:38} 
g_1=\frac{4}{3}g{_{A}^q}, \hspace{1cm} g_2=\sqrt{\frac{2}{3}}g{_{A}^q}.
\end{equation}

A nucleon axial coupling $g{_{A}^N}$ = 1.25 can be reproduced by taking $g{_{A}^q}$ = 0.75 \cite{Yan1992,Cho1994,Pirjol1997}. Therefore, the above equations are,
 
\begin{equation}
\label{eqn:39} 
g_1=1, \hspace{1cm} g_2=0.61.
\end{equation} 

Hence, the quark model predictions of coupling constants are in accordance with the results obtained in the framework of HHChPT (see in Eq. (\ref{eqn:37})). The lattice QCD studies \cite{Detmold2012} give $g_1$ and $g_2$ slightly different as, 0.56 $\pm$ 0.13 and 0.41 $\pm$ 0.08, respectively. Refs. \cite{Guralnik1993,Jenkins1993} used leading order approximation of large $N_c$ limit and calculated $g_2$ = $g{_{A}^q}/\sqrt{2}$ = 0.88 and, recently the chiral structure model predict, $g_2 = 0.688{_{-0.035}^{+0.013}}$ \cite{Kawakami2018}, which are larger than the experimental values and the quark model predictions (see in Eq. (\ref{eqn:39})). Taking an experimental average value of $\mid g_2 \mid_{2018}$ = $0.566{_{-0.027}^{+0.013}}$ and assuming the validity of the quark model relations among different coupling constants (see in Eq. (\ref{eqn:38})) implies $g{_{A}^q}$ = $0.694{_{-0.025}^{+0.013}}$ and $\mid g_1 \mid_{2018}$ = $0.925{_{-0.063}^{+0.017}}$.\\

The decay widths for the $P$-wave transitions are listed in the Table \ref{tab:8}. The first column represents the decay widths corresponding to the masses obtained in the screened potential with $\mid g_2 \mid$ = $0.550{_{-0.027}^{+0.013}}$ and the second column used the masses from PDG-2018 \cite{Tanabashi2018} with the same $g_2$. The third and the fourth column used the masses from PDG-2014 and PDG-2006 in the framework of HHChPT as in Ref. \cite{Cheng2015} and Ref. \cite{Cheng2007} respectively. We compare our results with other theoretical predictions and the experimental measurements. Moreover, the ratio of the decay widths of $\Sigma_c(2520)$ to $\Sigma_c(2455)$ is $\sim 7$ and it will be same in the limit of heavy quark symmetry.\\

\subsubsection{$S$ and $D$-wave transitions}

\begin{table}
\caption{The strong $S$ and $D$-wave transitions (in MeV) between $p$ and $s$-wave baryons.}
\scalebox{1}{
\label{tab:8}       % Give a unique label
\begin{tabular}{cccccccccccccccccccc}
\hline\noalign{\smallskip}
Decay & HHChPT & HHChPT & \cite{Cheng20151} & \cite{Cheng2007} & \cite{Gandhi2018} & \cite{Ivanov1999} & \cite{Tawfiq1998} & \cite{Mizuk2005} & PDG\cite{Tanabashi2018} \\
& Present & PDG-2018 &&&&&&&\\
\noalign{\smallskip}\hline\noalign{\smallskip}
$\Sigma_c(2800)^{++} \rightarrow \Lambda{_{c}^+} \pi^+$ & 72.00 & 72.53 & input & input & 72.67 &  &  & & 75 \\
&${_{-14.96}^{+16.56}}$&${_{-14.76}^{+16.30}}$&&&&&&&${_{-13-11}^{+18+12}}$&\\
\noalign{\smallskip}
$\Sigma_c(2800)^{+} \rightarrow \Lambda{_{c}^+} \pi^0$ & 68.58 & 67.04 & input & input & 64.54 & &  & & 62 \\
&${_{-14.11}^{+15.65}}$&${_{-13.56}^{+16.38}}$&&&&&&&${_{-23-38}^{+37+52}}$&\\
\noalign{\smallskip}
$\Sigma_c(2800)^{0} \rightarrow \Lambda{_{c}^+} \pi^-$ & 73.83 & 73.10 & input & input & 71.11 & & & & 72 \\
&${_{-14.47}^{+16.05}}$&${_{-14.96}^{+16.56}}$&&&&&&&${_{-15}^{+22}}$&\\
\noalign{\smallskip}
$\Lambda{_{c}(2625)^+} \rightarrow \Sigma_c(2455)^{++} \pi^-$ & $\approx$ 0 & 0.040 & 0.028 & 0.029 & 0.0012 & 0.076 & 2.15 & 0.011\\
&&${_{-0.012}^{+0.014}}$&&&& $\pm$ 0.009 &&\\
\noalign{\smallskip}
$\Lambda{_{c}(2625)^+} \rightarrow \Sigma_c(2455)^{0} \pi^+$ & $\approx$ 0 & 0.041 & 0.029 & 0.029 & 0.0013 & 0.080 & 2.61 & 0.011 & $<$0.97 \\
&&${_{-0.012}^{+0.014}}$&&&& $\pm$ 0.009 &&\\
\noalign{\smallskip}
$\Lambda{_{c}(2625)^+} \rightarrow \Sigma_c(2455)^{+} \pi^0$ & $\approx$ 0 & 0.044 & 0.040 & 0.041 & 0.0025 & 0.095 & 1.73 & 0.011\\
&&${_{-0.012}^{+0.014}}$&&&& $\pm$ 0.012 &&\\
\noalign{\smallskip}
$\Sigma_c(3/2^-)^{++} \rightarrow \Lambda{_{c}^+} \pi^+$ & 64.68 & input & & & & & & & 75 \\
&${_{-19.12}^{+22.68}}$&&&&&&&&${_{-13-11}^{+18+12}}$&\\
\noalign{\smallskip}
$\Sigma_c(3/2^-)^{+} \rightarrow \Lambda{_{c}^+} \pi^0$ & 71.46 & 69.82 & & & & & & & 62 \\
&${_{-21.12}^{+25.06}}$&${_{-22.87}^{+37.50}}$&&&&&&&${_{-23-38}^{+37+52}}$&\\
\noalign{\smallskip}
$\Sigma_c(3/2^-)^{0} \rightarrow \Lambda{_{c}^+} \pi^-$ & 75.31 & 78.56 & & & & & & & 72 \\
&${_{-22.26}^{+26.41}}$&${_{-26.64}^{+32.38}}$&&&&&&&${_{-15}^{+22}}$&\\
\noalign{\smallskip}\hline
\end{tabular}}
{\tiny{Ref. \cite{Cheng20151} $\rightarrow$ HHChPT(2015) PDG-2014 and\\
Ref. \cite{Cheng2007} $\rightarrow$ HHChPT(2007) PDG-2006.}}
% Or use
%\vspace*{5cm}  % with the correct table height
\end{table}

The $S$ and $D$-wave couplings take place between $p$-wave and $s$-wave baryons. The couplings $h_2,...,h_7$ are dimensionless employed for the $S$-wave couplings and $h_8,...,h_{15}$ used for the $D$-wave couplings has a dimension $E^{-1}$. In HHChPT, the coupling constants $h_2$ and $h_8$ can be extracted from the decays $\Lambda_c(2595)^+ \rightarrow \Sigma_c(2455) \pi$ and $\Lambda_c(2625)^+ \rightarrow \Sigma_c(2455) \pi$ respectively. Experimentally, the mass difference $M(\Lambda_c(2595)^+)-M(\Sigma_c(2455)) \approx $ 139 MeV, therefore, such a decay is kinematically barely allowed. The CDF \cite{Aaltonen2011} measured the decay width $2.59 \pm 0.30 \pm 0.47$ MeV$/c^2$ of $\Lambda_c(2595)^+$ into $\Lambda{_{c}^+ \pi^+ \pi^-}$ decay mode, having $h{_{2}^2}$ = 0.36 $\pm$ 0.04 $\pm$ 0.07, that implies $h_2$ = 0.60 $\pm$ 0.07, which is close to $h_2$ = 0.57${_{-0.197}^{+0.322}}$ of Ref. \cite{Pirjol1997} and 0.63 $\pm$ 0.07 of Ref. \cite{Cheng2015}. The decays of isotriplet $\Sigma_c(2800)$ into $\Lambda{_{c}^+} \pi$ are governed by the coupling $h_3$. For the $S$-wave transition, the quark model relations of the couplings are\\

\begin{equation}
\label{eqn:40} 
\mid h_4 \mid = 2 \mid h_2 \mid, \hspace{1cm} \mid h_3 \mid = \frac{\sqrt{3}}{2} \mid h_4 \mid;
\end{equation} 

\noindent and for the $D$-wave transitions \cite{Pirjol1997,Cheng2007,Cheng2015}

\begin{equation}
\label{eqn:41} 
\mid h_8 \mid = \mid h_9 \mid = \mid h_{10} \mid, \hspace{1cm} \mid h_{11} \mid = \sqrt{2} \mid h_{10} \mid.
\end{equation} 

The coupling $h_{10}$ can be extracted from the decay of isotriplet $\Sigma_c({\frac{3}{2}}^-)^{++,+,0}$ into $\Lambda{_{c}^+} \pi$ end particles. Using the world average masses of isotriplet $\Sigma_c(2800)^{++,+,0}$ and their respective decay widths from Ref. \cite{Tanabashi2018}, we obtained $h_{10}$ as, 

\begin{equation}
\label{eqn:42} 
%\begin{aligned}
\mid h_{10} \mid = 0.999{_{-0.161}^{+0.162}} \times 10^{-3},  \hspace{0.5cm} 0.942{_{-0.819}^{+ 0.436}}  \times 10^{-3},  \hspace{0.5cm} 0.956{_{-0.055}^{+ 0.129}} \times 10^{-3}
%\end{aligned}
\end{equation}

\noindent in MeV$^{-1}$ respectively, which are larger than the estimation of naive dimensional analysis $ 0.4 \times 10^{-3} $ MeV$^{-1}$ and compatible with $0.85{_{-0.08}^{+0.11}} \times 10^{-3}$ of Ref. \cite{Cheng2015}. Using $h_{10}$, the couplings $h_8$ and $h_{11}$ can be determined from the quark model relations and they are involved in the decays of $\Lambda{_{c}(2625)^{+}} \rightarrow \Sigma_c(2455) \pi$ and $\Sigma_c({\frac{3}{2}}^-)^{++,+,0} \rightarrow \Sigma_c(2455) \pi / \Sigma_c(2520) \pi $ respectively. Using the couplings $h_3 = 1.049 \pm 0.107 $, $h_{10} = 0.999{_{-0.161}^{+0.162}} \times 10^{-3}$ MeV$^{-1}$ $ = h_8$ and $h_{11} = 1.413{_{-0.227}^{+0.229}} \times 10^{-3}$ MeV$^{-1}$, we calculate the decay rates for the $S$ and $D$-wave transitions (see in Table \ref{tab:8}). Our results are compatible with Refs. \cite{Cheng2007,Cheng2015} and other theoretical predictions.\\

Note that the $J^P$ values of $\Sigma_c(2800)$ are not yet confirmed experimentally. It may have either ${\frac{1}{2}}^-$ or ${\frac{3}{2}}^-$ total spin. So here in the framework of HHChPT we are using the same masses from PDG-2018 \cite{Tanabashi2018} in both the cases of total spin. By considering $J^P = {\frac{1}{2}}^-$, calculated decay widths of isotriplet $\Sigma_c(2800)$  are close to PDG-2018 \cite{Tanabashi2018} rather than the $J^P = {\frac{3}{2}}^-$. Hence, the $\Sigma_c(2800)$ are more likely to be $\Sigma_c(1/2^-)$. Using $h_{8}$ = $0.999{_{-0.161}^{+0.162}} \times 10^{-3}$ MeV$^{-1}$ the $S$-wave transitions of $\Lambda{_{c}(2625)^+}$ into isotriplet $\Sigma_c(2455)$ are calculated. For these decays our present calculated masses doesn$'$t provide such amount of phase space and the masses from PDG-2018 \cite{Tanabashi2018} gives the decay rates which are overestimated from Refs. \cite{Cheng2007,Cheng20151,Gandhi2018,Mizuk2005}. Such differences are because of the selection of coupling $h_{8}$, HHChPT-2007 \cite{Cheng2007} and HHChPT-2015 \cite{Cheng2015} are used $h_{8}$ $\leq$ $0.85{_{-0.08}^{+0.11}}$ $\times 10^{-3}$ MeV$^{-1}$, and the Ref. \cite{Gandhi2018} used $h_{8}$ = 0.4 $\times 10^{-3}$ MeV$^{-1}$ from \cite{Pirjol1997}.

\subsection{Electromagnetic Properties}

To probe the electromagnetic properties (magnetic moments, transition magnetic moments, form factors etc.) of baryons, the study of radiative decay processes are important. Such properties are used to expose the inner structures of the baryons. In this section, we study the magnetic moments and transition magnetic moments of the ground state nonstrange singly charmed baryons in the constituent quark model. For radiative decays, the transitions are taking place among the participating baryons by an exchange of massless photons. So it doesn$'$t contain phase space restrictions and that$'$s why some of the radiative decay modes are contributed significantly to the total decay rate.\\

\subsubsection{Magnetic Moments}

The magnetic moment of the baryon $(\mu_B)$ is purely the function of masses and spin of their internal quarks constitutions. It can be expressed in the form of expectation value \cite{Gandhi2019,Gandhi2018,Dhir2009},

\begin{equation}
\label{eqn:43} 
\mu_B = \sum_{q} \left\langle \Phi_{sf} \middle| \hat{\mu}{_q}_z \middle| \Phi_{sf} \right\rangle;  \hspace{0.5cm} q=u,d,c.
\end{equation}  

\noindent Here, $ \Phi_{sf} $ is the spin-flavor wave function of the participating baryon and $\hat{\mu}{_q}_z$ is the $z$-component of the magnetic moment of the individual quark given by, 

\begin{equation}
\label{eqn:44} 
\hat{\mu}{_q}_z = \frac{e_q}{2 m{_{q}^{eff}}} \cdot \hat{\sigma}_{q_z},
\end{equation}  

\begin{table}
%\begin{center}
\caption{\label{tab:9}Magnetic moments of the nonstrange singly charmed baryons (in $\mu_N$).}
\scalebox{1}{
\begin{tabular}{cccccccccccccccccccccc}
\hline\noalign{\smallskip}
Baryon & Expression & Present & \cite{Sharma2010} & \cite{Wang2019} & \cite{Yang2018} & \cite{Bernotas2013}  & \cite{Patel2008} & \cite{Dhir2009} & Others \\
\noalign{\smallskip}\hline\noalign{\smallskip}
$ \Lambda{_{c}^{+}} $ & $\mu_c$ & 0.421 & 0.39 & & & 0.411 & 0.385 & 0.370 & 0.232 \cite{Shi2018} \\
&&&&&&&&&0.42 \cite{Faessler2006}\\
\noalign{\smallskip}
$ \Sigma_c(2455)^{++} $ &$\frac{4}{3}\mu_u - \frac{1}{3}\mu_c$ & 1.836 & 2.540 & 1.50 & 2.15 & 1.679 & 2.279 & 2.09 &  2.147 \cite{HCKim2019}\\
&&&& $\pm$ 0.32 & $\pm$ 0.1 &&&& 1.604 \cite{Shi2018} &\\
&&&&&&&&& 1.76 \cite{Faessler2006}&\\
&&&&&&&&& 2.220 \cite{Can2014}&\\
\noalign{\smallskip}
$ \Sigma_c(2455)^+ $ & $\frac{2}{3}\mu_u + \frac{2}{3}\mu_d - \frac{1}{3}\mu_c$ & 0.379 & 0.540 & 1.26 & 0.46 & 0.318 & 0.500 & 0.550 & 0.537  \cite{HCKim2019} \\
&&&& $\pm$ 0.09 & $\pm$ 0.03 &&&& 0.36 \cite{Faessler2006} &\\
&&&&&&&&& 1.00 \cite{Can2014} &\\
\noalign{\smallskip}
$ \Sigma_c(2455)^0 $ & $\frac{4}{3}\mu_d - \frac{1}{3}\mu_c$ & $-$1.085 & $-$1.46 & $-$0.97 & $-$1.24 & $-$1.043 & $-$1.015 & $-$1.230 & $-$1.073   \cite{HCKim2019} &\\
&&&& $\pm$ 0.14 & $\pm$ 0.05 &&&& $-$1.403  \cite{Shi2018} &\\
&&&&&&&&& $-$1.04  \cite{Faessler2006} &\\
&&&&&&&&& $-$1.073 \cite{Can2014} &\\
\noalign{\smallskip}
$ \Sigma_c(2520)^{++} $ & $ 2 \mu_u + \mu_c $ & 3.255 & 4.390 & 2.56 & 3.22 & 3.127 & 3.844 & 3.630 & 2.91 \cite{Meng2018}\\
&&&& $\pm$ 0.46 & $\pm$ 0.15 &&&&4.81&\\
&&&&&&&&&$\pm$ 1.22 \cite{Aliev2009}&\\
\noalign{\smallskip}
$ \Sigma_c(2520)^+ $ & $ \mu_u + \mu_d + \mu_c $ & 1.127 & 1.390 & 0.71& 0.68 & 1.085 & 1.256 & 1.180 & 0.99 \cite{Meng2018}\\
&&&& $\pm$ 0.13 & $\pm$ 0.04 &&&& 2.00 &\\
&&&&&&&&& $\pm$ 0.46 \cite{Aliev2009} &\\
\noalign{\smallskip}
$ \Sigma_c(2520)^0 $ & $ 2 \mu_d + \mu_c $ & $-$1.012 & $-$1.610 & $-$1.14 & $-$1.86 & $-$0.958 & $-$0.850 & $-$1.180 & $-$0.92 \cite{Meng2018} \\
&&&& $\pm$ 0.20 & $\pm$ 0.07 &&&& $-$0.81 &\\
&&&&&&&&& $\pm$ 0.20  \cite{Aliev2009} &\\
\noalign{\smallskip}\hline
\end{tabular}}
%\end{center}
\end{table}

\noindent where $e_q$ is the charge and $\hat{\sigma}_{q_z}$ is the $z$-component of the spin of the constituent quark. The effective quark mass $(m{_{q}^{eff}})$ gives the mass of the bound quark inside the baryon by taking into account its binding interactions with other two quarks and is defined as,

\begin{equation}
\label{eqn:45} 
m{_{q}^{eff}} = m{_{q}} \left({1 + \frac{\left\langle H \right\rangle}{\sum\limits_{q} m{_{q}}}}\right),  
\end{equation}

\noindent where $\sum\limits_{q} m_{q}$ is the sum of constituting quark mass and the Hamiltonian $ {\left\langle  H \right\rangle} = M - {\sum\limits_{q} m{_{q}}} $, where $M$ is the measured or predicted baryon mass.\\

Using these equations and taking the constituent quark masses from section 2 and the baryon masses from the spectrum, we determine the ground state magnetic moments of the nonstrange singly charmed baryons in the unit of nuclear magnetons $\left(\mu_N = \frac{e\hbar}{2m_p}\right)$. We present our results with the predictions obtained from various approaches in Table \ref{tab:9}.\\

\subsubsection{Radiative Decays}

An expression of electromagnetic decay width is written in the form of radiative transition magnetic moments $(\mu_{B_{C}^{\prime}\rightarrow B_{c}})$ \cite{Gandhi2018,Majethiya2009},

\begin{equation}
\label{eqn:46} 
\Gamma_{\gamma} = \frac{k^3}{4\pi}\frac{2}{2J+1}\frac{e}{m{_{p}^2}}\mu{_{{B{_{c}}} \rightarrow {B{_{c}^{\prime}}}}^2}.
\end{equation}

\noindent Here, k is the photon momentum,
 
\begin{equation}
\label{eqn:47} 
k = \frac{M{_{B_c}^{2}} - M{_{B_c^{\prime}}^{2}}}{2M_{B}{_c}},
\end{equation}

\noindent $ m_p $ is the mass of proton and $ J $ represents the total angular momentum of the initial baryon $({B{_{c}}})$. $M{_{B_c}}$ and $M{_{B_c^{\prime}}}$ are the mass of the initial and final state baryon respectively. \\

\begin{table}
\caption{\label{tab:10}The transition magnetic moments $ \left\vert \mu{_{{B_c} \rightarrow B{_{c}^{\prime}}}} \right\vert $ of nonstrange singly charmed baryons (in $\mu_N $).}
\begin{tabular}{ccccccccccccccccccc}
\hline\noalign{\smallskip}
Transition & Expression & Present & \cite{Wang2019} & \cite{Aliev20091} & \multicolumn{3}{c}{\cite{Dhir2009} } & \cite{Majethiya2009}\\
\noalign{\smallskip}
\cline{6-8}
\noalign{\smallskip}
&&& & & (nqm) & (ems) & (ses)  &  & & \\
\noalign{\smallskip}\hline\noalign{\smallskip}
$ \mu_{\Sigma_c(2455)^+ \rightarrow \Lambda{_{c}^{+}}} $ & $ \frac{-1}{\sqrt{3}} (\mu_u - \mu_d) $  & 1.266 &  & 1.48 & 2.28 & 2.28 & 2.15 & 1.347 & \\
&&&&$\pm$ 0.55&\\
\noalign{\smallskip}
$ \mu_{\Sigma_c(2520)^{++} \rightarrow \Sigma_c(2455)^{++}} $ & $ \frac{2\sqrt{2}}{3} (\mu_u - \mu_c) $ & 0.996 & 1.07 & 1.06 & 1.41 & 1.19 & 1.23 & 1.080 & \\
&&&$\pm$ 0.23&$\pm$ 0.38&\\
\noalign{\smallskip}
$ \mu_{\Sigma_c(2520)^{+} \rightarrow \Sigma_c(2455)^{+}} $ & $ \frac{\sqrt{2}}{3} (\mu_u + \mu_d - 2\mu_c) $ & 0.009 & 0.19 & 0.45 & 0.09 & 0.04 & 0.08 & 0.008 & \\
&&&$\pm$ 0.06&$\pm$ 0.11&\\
\noalign{\smallskip}
$ \mu_{\Sigma_c(2520)^{0} \rightarrow \Sigma_c(2455)^{0}} $ & $ \frac{2\sqrt{2}}{3} (\mu_d - \mu_c) $ & 1.012 & 0.69 & 0.19 & 1.22 & 1.11 & 1.07 & 1.064 &  \\
&&&$\pm$ 0.1&$\pm$ 0.08&\\
\noalign{\smallskip}
$ \mu_{\Sigma_c(2520)^{+} \rightarrow \Lambda{_{c}^{+}}} $ & $ \sqrt{\frac{2}{3}} (\mu_u - \mu_d) $  & 1.744 & &  & & & & 1.857 & \\
\noalign{\smallskip}\hline
\end{tabular}
\end{table}

\begin{table*}
\begin{center}
\caption{\label{tab:11}The radiative decay widths ($ \Gamma_\gamma $) of nonstrange singly charmed baryons (in keV).}
\begin{tabular}{ccccccccccccccccccccccc}
\hline\noalign{\smallskip}
Decay & Present & \cite{Cheng1997} & \cite{Jiang2015} & \cite{Wang2019} & \cite{Ivanov1999} & \cite{Aliev2012} & \cite{Bernotas2013} & \cite{Wang2017} & \cite{Majethiya2009} \\
\noalign{\smallskip}\hline\noalign{\smallskip}
$\Sigma_c(2455)^+$ $ \rightarrow \Lambda{_{c}^{+}} \gamma $ & 71.20 & 88 & 164 & 65.6 & 60.7 & & & 80.6 & 60.55 \\
&&&&$\pm$ 2&$\pm$ 1.5&&&&\\
\noalign{\smallskip}
$\Sigma_c(2520)^{++}$ $ \rightarrow \Sigma_c(2455)^{++} \gamma $ & 1.315 & 1.4 & 11.6 & 1.20 &  & 3.567 & 0.826 & 3.94 & 1.15 \\
&&&&$\pm$ 0.6&&&&&\\
\noalign{\smallskip}
$\Sigma_c(2520)^{+}$  $ \rightarrow \Sigma_c(2455)^{+} \gamma $ & 1 $\times$ 10$^{-4}$ & 0.002 & 0.85 & 0.04 & 0.14 & 0.187 & 0.004 & 0.004 & 6 $\times$ 10$^{-5}$ \\
&&&&$\pm$ 0.03&$\pm$ 0.004&&&&\\
\noalign{\smallskip}
$\Sigma_c(2520)^{0}$ $ \rightarrow \Sigma_c(2455)^{0} \gamma $ & 1.072 & 1.2 & 2.92 & 0.49 &  & 1.049 & 1.08 & 3.43 & 1.12\\
&&&&$\pm$ 0.1&&&&&\\
\noalign{\smallskip}
$\Sigma_c(2520)^{+}$ $ \rightarrow \Lambda{_{c}^{+}} \gamma $ & 171.9 & 147 & 893 & 161.6 & 151 & 409.8 & 126 & 373 & 154.48\\
&&&&$\pm$ 5&$\pm$ 4&&&&\\
\noalign{\smallskip}\hline
\end{tabular}
\end{center}
\end{table*}

For transition magnetic moments $ (\mu_{B_c \rightarrow B{_{c}^{\prime}}}) $, repeating the same procedure as we discussed in the above subsection by sandwiching the magnetic moment operator (Eq. (\ref{eqn:44})) between the appropriate initial $ (\Phi_{sf_{B{_{c}}}}) $ and final $ (\Phi_{sf_{B{_{c}^{\prime}}}}) $ state spin-flavour wave functions of nonstrange singly charmed baryons,

\begin{equation}
\label{eqn:48} 
\mu{_{{B_c} \rightarrow B{_{c}^{\prime}}}} = \langle\Phi_{sf_{B{_{c}}}} \vert \hat{\mu}{_q}_z \vert \Phi_{sf_{B{_{c}^{\prime}}}} \rangle
\end{equation}

\noindent For example: in order to determine the radiative decay of $\Sigma_c(2520)^{++}$ into $\Sigma_c(2455)^{++}$, first we need to calculate the transition magnetic moment as,
 
\begin{equation}
\label{eqn:49} 
\mu_{\Sigma_c(2520)^{++} \rightarrow \Sigma_c(2455)^{++}} = \left \langle\Phi_{sf_{\Sigma{_{c}(2520)^{++}}}} \right \vert \hat{\mu}{_q}_z \left\vert \Phi_{sf_{\Sigma{_{c}(2455)^{++}}}}\right\rangle
\end{equation}

\noindent the spin-flavour wave functions $ (\Phi_{sf}) $ of $\Sigma_c(2520)^{++}$ and $\Sigma_c(2455)^{++}$ baryons are expressed as,

\begin{equation}
\label{eqn:50} 
\left\vert \Phi_{sf_{\Sigma_c(2520)^{++}}} \right\rangle  = (uuc) \cdot \left(\frac{1}{\sqrt{3}}(\uparrow\uparrow\downarrow+\uparrow\downarrow\uparrow+\downarrow\uparrow\uparrow)\right)
\end{equation}

\begin{equation}
\label{eqn:51} 
\left\vert \Phi_{sf_{\Sigma_c(2455)^{++}}} \right\rangle  = (uuc) \cdot \left(\frac{1}{\sqrt{6}}(2\uparrow\uparrow\downarrow-\uparrow\downarrow\uparrow-\downarrow\uparrow\uparrow)\right)
\end{equation}

\noindent Following the orthogonal condition of quark flavor and spin states, for example $ \left\langle u\uparrow u \uparrow c\downarrow \vert u \uparrow u\downarrow c\uparrow \right\rangle = 0 $, we get an expression of transition magnetic moment as

\begin{equation}
\label{eqn:52} 
\mu_{\Sigma_c(2520)^{++} \rightarrow \Sigma_c(2455)^{++}} = \frac{2\sqrt{2}}{3} \left (\mu_u - \mu_c \right).  
\end{equation}

\noindent In this way, we determine an expressions of transition magnetic moments of other nonstrange singly charmed baryons (see in Table \ref{tab:10}). Our calculated transition magnetic moments and their corresponding radiative decay widths are listed in Tables \ref{tab:10} and \ref{tab:11} respectively. Our results are compared with other theoretical predictions and are in accordance with A. Majethiya \textit{et al.} \cite{Majethiya2009} and smaller than the predictions of Ref. \cite{Dhir2009}.

\section{Summary}
\label{sec:5}

In this work, the excited state mass spectra of nonstrange singly charmed baryons are calculated in the framework of hypercentral Constituent Quark Model (hCQM). Here, we have used screened potential as a confining potential. In order to see the relativistic effect in the heavy light baryonic systems we added the first order correction. Our calculated masses are listed in Tables \ref{tab:2}-\ref{tab:6}. The predicted mass spectra are in agreement with the other theoretical predictions and the experimental measurements where available. For the ${\Lambda{_{c}^+}}$ baryon, our results are in consistent with the results obtained by Z. Shah \textit{et al.} \cite{Shah2016epja} using the linear confinement potential as a scalar potential. We have seen the screening effect in isotriplet $\Sigma_c$ baryons, which gives lowered masses for higher excited states compared to the masses obtained from the linear potential. The calculations of excited states masses allow us for plotting the data on Regge line according to their quantum number with natural and unnatural parities in $(J, M^2)$ plane, which helps to assign the $J^P$ value of an experimental unknown states. Regge trajectories identified the $ {\Lambda{_{c}(2765)^+}} $ baryon with $2S$ state and having a spin $\frac{1}{2}$ in our case (see in Fig. (\ref{Fig:2})). Because of the screening effect in isotriplet ${\Sigma{_{c}^+}}$ baryon the Regge trajectories approaches a straight horizontal line for the higher excited states as shown in Fig. (\ref{Fig:3}). That are in accordance with the argument of decreasing the Regge slopes with increasing the quark masses (see Refs. \cite{Guo2008,Zhang2005,Li2004,Brisudova2000}). From the Fig. (\ref{Fig:3}) we can not predict exactly the spin-parity of the first orbital excited state of isotriplet $\Sigma_c$ baryons.\\        

The strong decay rates of nonstrange singly charmed baryons are analyzed in the framework of Heavy Hadron Chiral Perturbation Theory (HHChPT). Using the masses and the decay widths of these baryons from PDG-2018 \cite{Tanabashi2018}, first we have extracted couplings constants $g_2$, $h_2$ (from CDF \cite{Aaltonen2011}) and $h_{10}$ from their respective decay channels. And the couplings $h_3$, $h_8$, $h_9$ and $h_{11}$ are obtained from the quark model relations. Such couplings are in accordance with the quark model expectation and other theoretical predictions. These coupling control the strong one pion decay rates in HHChPT. Using the masses from PDG-2018 \cite{Tanabashi2018} and from the screened potential spectrum the strong decay rates for the $S$, $P$ and $D$-wave transitions are calculated separately. Our results are listed in Tables \ref{tab:7} and \ref{tab:8}, and compared with other predictions. The strong decay rates of isotriplet $\Sigma_c(2800)$ baryons are calculated with the total spin ${\frac{1}{2}}^-$ and ${\frac{3}{2}}^-$. Their decay rates corresponding to spin ${\frac{1}{2}}^-$ are close to the experimental observations rather than the spin ${\frac{3}{2}}^-$. Hence, we assign the $J^P$ quantum number of isotriplet $\Sigma_c(2800)$ as ${\frac{1}{2}}^-$. The magnetic moments, transition magnetic moments and the radiative decay rates are calculated for the ground state nonstrange singly charmed baryons in the constituent quark model, which are presented in Tables \ref{tab:9}, \ref{tab:10} and \ref{tab:11} with other theoretical predictions.\\

In the present study, our aim is satisfied for the determination of $J^P$ value of experimentally measured unknown states of nonstrange singly charmed baryons: $\Lambda{_{c}(2765)^+}$ and isotriplet $\Sigma_c(2800)$. The spectroscopy and the Regge trajectories predict the $\Lambda{_{c}(2765)^+}$ as a $2S$ state with $J^P$ = ${\frac{1}{2}}^+$. And the strong decays analysis of isotriplet $\Sigma_c(2800)$ baryons predict $J^P$ = ${\frac{1}{2}}^-$ as a first orbital excitation. This model is successful for the study of nonstrange singly charmed baryons. Our predictions will help the experimentalists as well as theoreticians towards the understanding of their dynamics. So we would like to extend this scheme for the study of singly bottom baryons.

\end{document}